\documentclass[aps,amssymb,10pt,showpacs,showkeys,letterpaper]{revtex4}
\usepackage[utf8]{inputenc}
\usepackage{graphicx}
\usepackage{amsmath}
\usepackage{epsfig}
\usepackage{bm}

\usepackage{enumitem}
\usepackage{dcolumn}
\usepackage{array,multirow}
\usepackage{epstopdf}
\usepackage{amsfonts}
\usepackage{amssymb}
\usepackage{tabularx}

\usepackage{color}
\usepackage{xcolor}

\newcommand{\be}{\begin{equation}}
\newcommand{\ee}{\end{equation}}
\newcommand{\beq}{\begin{eqnarray}}
\newcommand{\eeq}{\end{eqnarray}}

\begin{document}

\title{Massive Dirac quasinormal modes in Schwarzschild-dS black holes:\\
anomalous decay rate and fine structure.}

\author{Almendra Arag\'{o}n}
\email{almendra.aragon@mail.udp.cl} \affiliation{Facultad de
Ingenier\'{i}a y Ciencias, Universidad Diego Portales, Avenida Ej\'{e}rcito
Libertador 441, Casilla 298-V, Santiago, Chile.} 

\author{Ram\'{o}n B\'{e}car}
\email{rbecar@uct.cl}
\affiliation{Departamento de Ciencias Matem\'{a}ticas y F\'{\i}sicas, Universidad Catolica de Temuco}

\author{P. A. Gonz\'{a}lez}
\email{pablo.gonzalez@udp.cl} \affiliation{Facultad de
Ingenier\'{i}a y Ciencias, Universidad Diego Portales, Avenida Ej\'{e}rcito
Libertador 441, Casilla 298-V, Santiago, Chile.}

\author{Yerko V\'asquez}
\email{yvasquez@userena.cl}
\affiliation{Departamento de F\'isica, Facultad de Ciencias, Universidad de La Serena,\\
Avenida Cisternas 1200, La Serena, Chile.}

\date{\today}

\begin{abstract}

Recently, the anomalous decay rate of quasinormal modes has been studied for some geometries under scalar field perturbations, which occurs when the longest-lived modes are the ones with higher angular number, as well as, the existence of a critical scalar field mass, i.e the value of scalar field mass such that the decay rate does not depend appreciably on the angular number, and beyond which
the behaviour of the decay rate is inverted. Here, we consider the propagation of fermionic fields in the background of Schwarzschild de Sitter black holes, and we show that the  anomalous decay rate behaviour and the fine structure, related to the coupling between the chirality and the mass of the field, can be observed in the fermionic spectrum.

\end{abstract}

\maketitle

\tableofcontents


\newpage
\section{Introduction}

Quasinormal modes (QNMs) 
and the frequencies of QNMs or quasinormal frequencies (QNFs) have been a subject of study for a long time 
\cite{Regge:1957td, Zerilli:1971wd, Kokkotas:1999bd, Nollert:1999ji, Konoplya:2011qq} and  have recently acquired great interest due  
to the detection 
of gravitational waves \cite{Abbott:2016blz,TheLIGOScientific:2016src}.
The QNMs of different spin fields on
Schwarzschild,  Schwarzschild-de Sitter and Schwarzschild anti-de Sitter spacetimes have been studied extensively, being the scalar field the most studied case. Considering the scenario when the black hole is immersed in an expanding universe, the QNMs of black holes in de Sitter (dS) space result of interest. 
However, due to the observed value of the cosmological constant
is very small, 
it would
be reasonable to ask why we should not neglect its effects 
in local physics, or more precisely, how strong are its perturbative effects.
Also, from a modern point of view, the study of the dynamical properties of dS black holes provides important insight into de Sitter/conformal field theory correspondence (dS/CFT) \cite{Strominger:2001pn,Abdalla:2002rm,Abdalla:2002hg}.
Thus, following these interesting issues it is natural to study the propagation of test field in dS spacetimes.

The gravitational QNMs of Schwarzschild-de Sitter black hole were studied 
in Ref. \cite{ Mellor:1989ac} by following a procedure analogous to that of Chandrasekhar \cite{Chandra:1983}, and based on the Wentzel-Kramers-Brillouin (WKB) approximation in Ref. \cite{Otsuki}. Then, the existence of exponentially decaying tails at late times was demonstrated in Ref. \cite{Brady:1996za} and the dynamics of radiative fields was analyzed in Ref. \cite{Brady:1999wd}. After that,  via an analytical expression for the QNMs and QNFs of nearly extreme Schwarzschild–de Sitter black holes, it was demonstrated
that 
the P\"oschl-Teller is the true potential \cite{Cardoso:2003sw}, 
in good agreement with those found by Moos et al \cite{Moss:2001ga}. Other analytical approximation to the problem of scalar field perturbation on Schwarzschild-de Sitter black holes was studied, and it was shown the presence of two sets of modes relevant at two different time scales, proportional to the surface gravity of the black hole and to the cosmological horizon respectively  \cite{Suneeta:2003bj}. As well, the QNMs were calculated by using the sixth order WKB formula and the approximation by the P\"oschl--Teller potential, and it was shown that the QNFs all have a negative imaginary part, which means that the propagation of scalar field is stable in this background, and the presence of the cosmological constant leads to decrease of the frequency of oscillation and to a slower decay rate \cite{Zhidenko:2003wq}. Further, high overtones of gravitational and electromagnetic QNMs were studied in Ref. \cite{Konoplya:2004uk}. Also, an exhaustive analysis was performed by considering the WKB approach and two numerical schemes: the characteristic and general initial value integrations \cite{Molina:2003dc}. Additionally, a
simple derivation of the imaginary parts of the QNFs 
was proposed by calculating the scattering amplitude in the first Born approximation and determining its poles \cite{Choudhury:2003wd}. Later, the QNMs of massless scalar field was studied for a  Reissner-Nordstr\"om de Sitter (RNdS) black hole with a global monopole by the 6th order WKB approximation, and it was discussed in detail how the parameters of black hole space-time influence the QNMs of massless scalar field \cite{Zhang:2014xha}. Also, the phenomenon of slowly decaying resonances for Schwarzschild-de Sitter spacetimes, in the large scalar mass approximation was studied in Ref.  \cite{Toshmatov:2017qrq}. Besides, it was found a novel infinite set of purely imaginary modes, which depending on the black hole mass may even be the dominant mode  \cite{Jansen:2017oag}, and the different families of modes were recently study in detail in Ref. \cite{Aragon:2020tvq},  and the different families of massless scalar fields in the exterior of RNdS black   holes were analyzed in Ref. \cite{Cardoso:2017soq}. Finally, very recently it was investigated the relaxation rate of the d- dimensional  RNdS black hole perturbed by neutral massless scalar field in the eikonal limit \cite{Zhang:2020zic}, finding that fastest rate relaxation increases with the cosmological constant for all dimensions, in addition it was analyzed the relationship between the cosmological constant and the critical charge and their influence on  relaxation rate of the composed system.

On the other hand, it was recently shown that the decay rate of scalar QNMs in Schwarzschild, Schwarzschild-de Sitter and Schwarzschild-AdS black holes present an anomalous behaviour, i.e the longest-lived modes are the ones with higher angular number, while that for Schwarzschild and Schwarzschild-de Sitter there is a critical scalar field mass, such that beyond this value the anomalous decay rate behaviour is inverted \cite{Lagos:2020oek,Aragon:2020tvq}. Also, such anomalous behaviour was shown for black holes in $f(R)$ gravity \cite{Aragon:2020xtm}, Reissner-Nordstr\"om black holes \cite{Fontana:2020syy},  and accelerating black holes \cite{Destounis:2020pjk}.

Having in mind that the studies of anomalous decay rate have been performed for scalar perturbations, in this work, we study the propagation of massive fermionic fields in Schwarzschild-de Sitter black hole backgrounds, in order to see,  if the anomalous decay rate behavior and 
a critical fermionic field mass are present in the spectrum, as well as the fine structure. We carry out this study by using the pseudospectral Chebyshev method \cite{Boyd}
which is an effective method to find high overtone modes \cite{Finazzo:2016psx,Gonzalez:2017shu,Gonzalez:2018xrq,Becar:2019hwk,Aragon:2020qdc}. It is worth mentioning that the study of the QNMs of the Dirac field has been performed for massless fermionic fields \cite{Zhidenko:2003wq,Jing:2003wq,Konoplya:2020zso} and massive fermionic fields \cite{Chang:2006kf}. The results show that the field with higher masses and larger cosmological constant will decay more slowly in a Schwarzschild-de Sitter black hole by using the WKB approximation \cite{Chang:2006kf}. Also, the Dirac QNMs of $D$-dimensional de Sitter spacetime was determined in Ref. \cite{LopezOrtega:2007sr}. On the other hand, it was shown that the two chiralities of massive fermions lead to an additional fine structure in the spectrum, for Schwarzschild and Kerr backgrounds  by using the convergent Frobenius method \cite{Konoplya:2017tvu}.

The manuscript is organized as follows: In Sec. \ref{QNM}, we give a brief review of the spacetime that we have considered, and we study the Dirac equation. Then, in Sec. \ref{numerical},
we calculate the QNFs of massless and massive fermionic fields numerically by using the pseudospectral Chebyshev method, and  for massive fields we study the fine structure and the anomalous decay rate. Finally, we conclude in Sec. \ref{conclusion}.

\section{Fermionic perturbations} 
\label{QNM}
The Schwarzschild de Sitter black holes
 are 
 solutions of the equations of motion that arise from the Einstein-Hilbert action with a positive cosmological constant
\begin{equation}
    S=\frac{1}{16\pi G}\int d^4x\sqrt{-g}(R-2\Lambda)\,,
\end{equation}
where $G$ is the Newton constant, $R$ is the Ricci scalar and $\Lambda$ is the cosmological constant. The Schwarzschild de Sitter black hole is described by the metric
\begin{equation}
    ds^2=f(r)dt^2-\frac{dr^2}{f(r)}-r^2 d\Omega^2 \,,
    \label{metric}
\end{equation}
where $d \Omega^2= d\theta^2+sin^2\theta d\phi^2$ , $f(r)=1-\frac{2M}{r}-\frac{\Lambda r^2}{3}$, $M$ is the black hole mass and $\Lambda>0$.  This metric represents a black hole when $\Lambda < 1/(9M^2)$, with two horizons: the event horizon $r_H$ and the cosmological horizon $r_{\Lambda}$, where $r_H < r_{\Lambda}$, and for $\Lambda= 1/(9M^2)$ both horizons coincide.

In order to study the propagation of fermionic fields in the background of 
Schwarzschild de Sitter black holes we consider the Dirac equation in curved space given by
\begin{equation}
\label{DE}
\left( \gamma ^{\mu }\nabla _{\mu }+m\right) \psi =0~,
\end{equation}%
where the covariant derivative is defined as 
$\nabla _{\mu }=\partial _{\mu }+\frac{1}{2}\omega _{\text{ \ }\mu
}^{ab}J_{ab}$,
with  $\omega^{ab}$ being the Levi-Civita spin connection and $J_{ab}=\frac{1}{4}\left[ \gamma _{a},\gamma _{b}\right]$ correspond to the generators of the Lorentz group, which are defined through the gamma matrices in a flat spacetime $\gamma ^{a}$, that can be expressed through 
the gamma matrices in curved space-time $\gamma ^{\mu }$ by  
$\gamma ^{\mu }=e_{\text{ \ }a}^{\mu }\gamma ^{a}$.
So, in order to
solve the Dirac equation (\ref{DE}), we use the diagonal vielbein
\begin{equation}
e^{0} = \sqrt{f} \, dt, \, \, \, e^{1} = \frac{1}{\sqrt{f}}dr,\, \, \, e^{2} = r \, d\theta, \, \, \, e^{3} = r \sin \theta \, d\phi \,,
\end{equation}
and from the null torsion condition 
$de^{a}+\omega _{\text{ \ }b}^{a}\wedge e^{b}=0$,
we obtain the nonzero components of the spin connection 
\begin{equation}
    \omega ^{01} = \frac{f'(r)}{2} \, dt, \,\, \omega ^{12} = -\sqrt{f} \, d\theta, \,\, \omega ^{13} = -\sqrt{f} \sin \theta \,d\phi, \,\,
    \omega ^{23} = -\cos \theta \, d\phi \,.
\end{equation}
Now, using the following representation of the gamma matrices \cite{Cho:2003qe,Du:2004jt}
$\gamma ^{0}=i\sigma ^{2}\otimes \mathbf{1}~,\text{ \ }\gamma ^{1}=\sigma
^{1}\otimes \mathbf{1}~,\text{ \ }\gamma ^{m}=\sigma ^{3}\otimes \tilde{%
\gamma}^{m}$,
where $\sigma ^{i}$ are the Pauli matrices, and $\tilde{\gamma}^{m}$ are the
Dirac matrices in the base manifold $\Omega$, along with the
following ansatz for the fermionic field 
\begin{equation}
\psi =\frac{e^{-i\omega t}}{rf^{1/4} }\left( 
\begin{array}{c}
\psi _{1} \\ 
\psi _{2}%
\end{array}%
\right) \otimes \varsigma ~,
\end{equation}%
where $\psi_1$ and $\psi_2$ are functions of $r$ and $\varsigma(\theta, \phi)$ is a $2$-components fermion, we obtain the following equations:
\begin{eqnarray}
\label{diff}
\notag -\frac{i\omega}{\sqrt{f}}\psi _{2}+\sqrt{f}\psi'_{2}+\frac{i  \kappa}{r}\psi _{1}+m\psi _{1} &=&0~,   \\ 
\frac{i\omega}{\sqrt{f}}\psi _{1}+\sqrt{f}\psi'_{1}-\frac{i \kappa}{r}\psi _{2}+m\psi _{2} &=&0~,  
\end{eqnarray}%
where $i \kappa$ is the eigenvalue of the Dirac operator on the $2$-dimensional sphere, and $\kappa$ can take positive and negative integer values $\kappa= \pm (\ell+1)$, with $\ell=0,1,2, ...$, and the prime denotes the derivative with respect to the radial coordinate $r$. These equations can be decoupled as
\begin{equation} \label{q1}
\psi_1''+\left(\frac{1}{2}\frac{f'(r)}{ f(r)}+\frac{i \kappa }{r (i \kappa -m r)}\right)\psi_1'+\frac{r^2 \omega  (m r- i \kappa ) \left(2 \omega -i f'(r)\right)-2 f(r) \left((i \kappa +m r) (i \kappa -m r)^2- \kappa  r \omega \right)}{2 r^2 f(r)^2 (m r-i \kappa )} \psi_1=0\,,
\end{equation}
\begin{equation} \label{q2}
\psi_2''+\left(\frac{1}{2}\frac{f'(r)}{ f(r)}+\frac{i \kappa }{r ( i \kappa +m r)}\right)\psi_2'+\frac{r^2 \omega  (m r+ i \kappa ) \left(2\omega +i f'(r)\right)-2 f(r) \left((-i \kappa +m r) ( i \kappa +m r)^2- \kappa  r \omega \right)}{2 r^2 f(r)^2 (m r+ i \kappa )} \psi_2=0\,.
\end{equation}
Notice that (\ref{q2}) can be obtained from (\ref{q1}) by means of the substitutions $\psi_1 \rightarrow \psi_2$, $\omega \rightarrow -\omega$ and $\kappa \rightarrow -\kappa$. On the other hand, the above equations (\ref{diff}), for a massless fermionic field, can be reduced to \begin{eqnarray}
\label{a1}&& -i \omega Z_{+}-\frac{dZ_{-}}{d r^{\ast}}= W Z_{-}~,  \\
\label{a2}&& -i \omega Z_{-}-\frac{dZ_{+}}{dr^{\ast}}= -W Z_{+}~,
\end{eqnarray}
where we have defined $Z_{\pm}=\psi_1 \pm i \psi_2$ and $W=-i \kappa \sqrt{f}/r $, see \cite{Chandra:1983}, and the tortoise coordinate $r^{\ast}$ is defined as usual by $d r^{\ast}=dr/f$. Now, decoupling (\ref{a1}) and (\ref{a2}), we obtain the following Schr\"odinger-like equations:
\begin{equation} \label{Schrodinger}
-\frac{d^2Z_\pm}{d r^{\ast 2}}+V_\pm=\omega^2 Z_\pm~, 
\end{equation}
where the effective potentials $V_{\pm}$ are given by
\begin{equation}
V_\pm=W^2 \pm \frac{dW}{d r^{\ast}} = \mp \kappa \frac{f \sqrt{f}}{r^2} \pm \kappa \frac{f' \sqrt{f}}{2r}+ \frac{\kappa^2 f}{r^2}~.
\end{equation}
We can observe that the potentials are not positive-definite. Also it is possible to demonstrate that the effective potentials for massive fermionic field are
\begin{equation}
\label{Pott}
V_\pm=W^2 \pm \frac{dW}{d r^{\ast}} \,,
\end{equation}
where
\begin{equation}
W[r]=\frac{\sqrt{f(r)}\sqrt{m^2+\frac{\kappa^2}{r^2}}}{1+
\frac{1}{2w}f(r)(\frac{m\kappa}{m^2 r^2+\kappa^2})} \,.
\end{equation}

\begin{figure}[h]
\begin{center}
\includegraphics[width=0.46\textwidth]{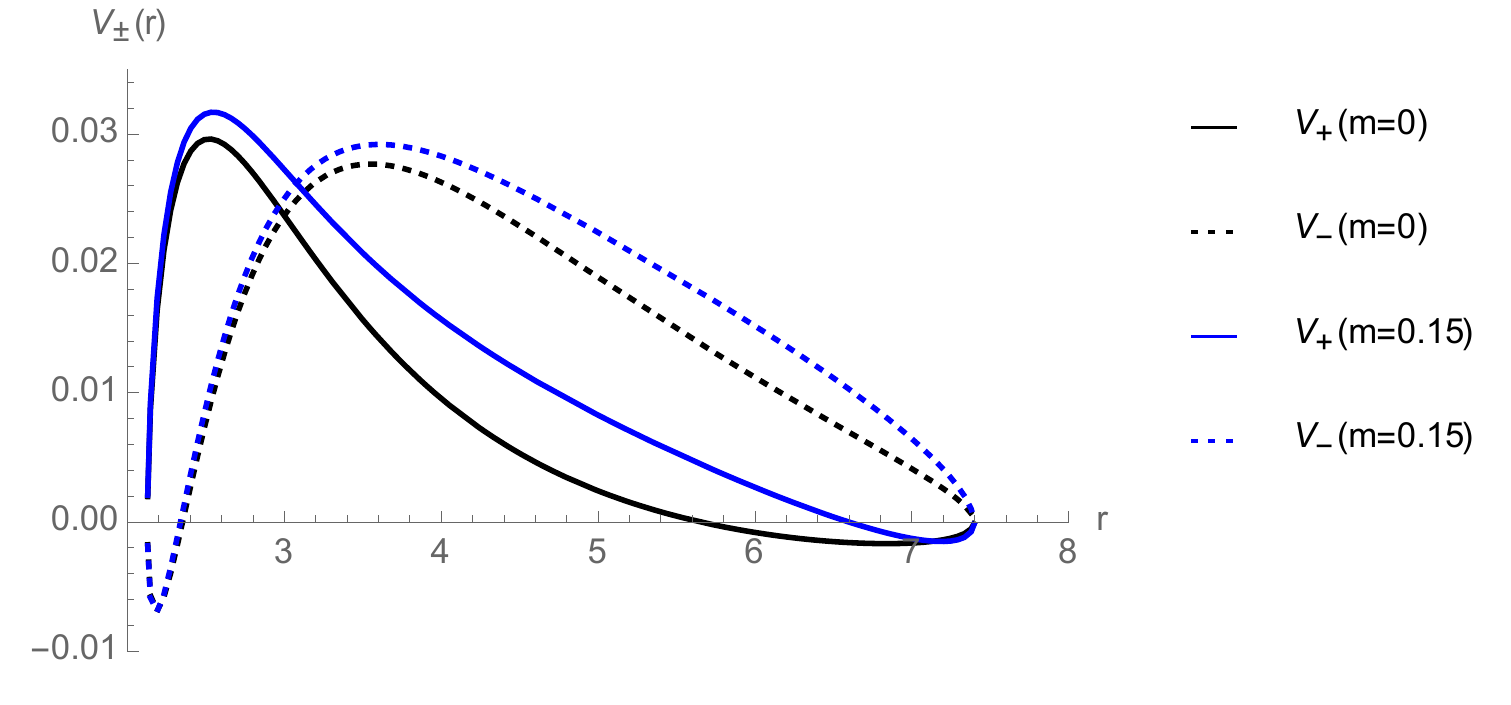}
\includegraphics[width=0.46\textwidth]{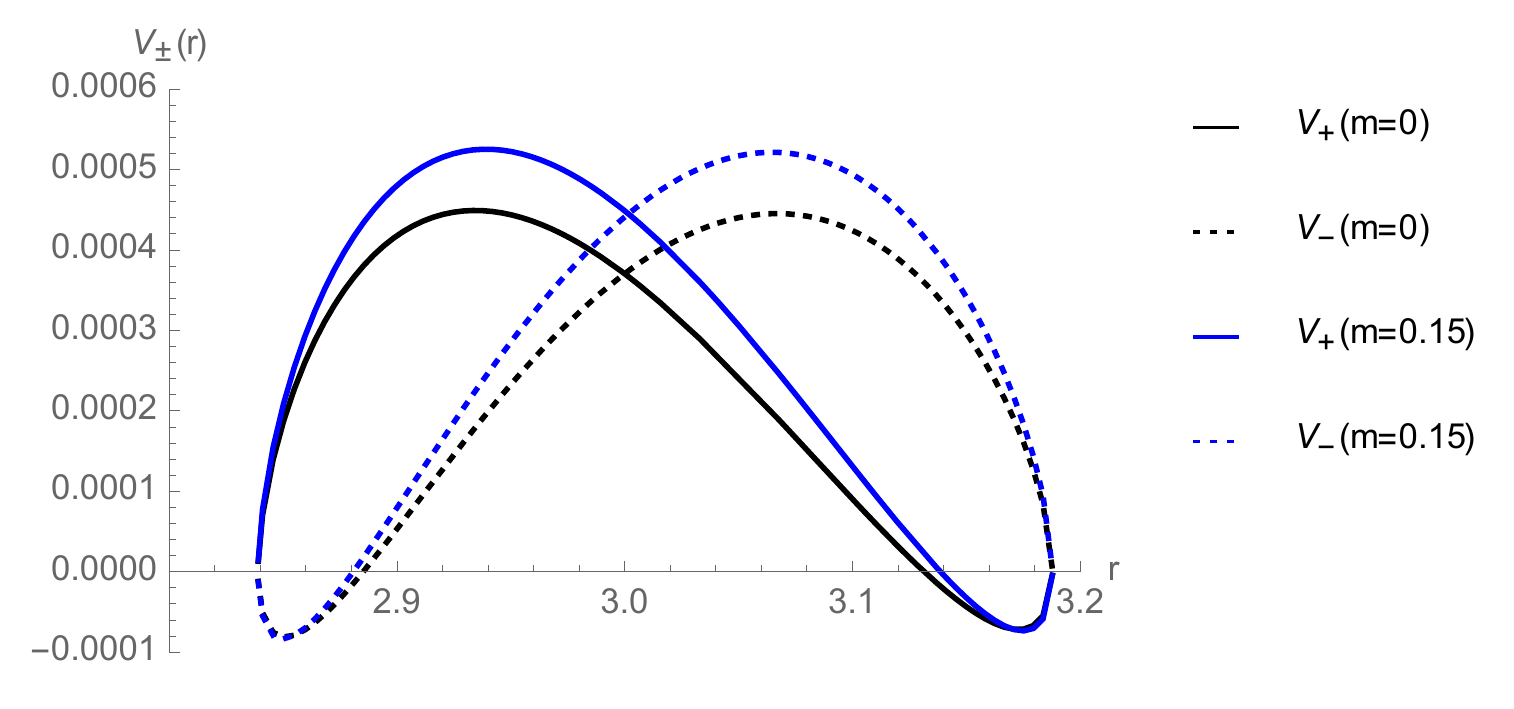}
\end{center}
\caption{The behaviour of $V_{\pm}(r)$ as a function of $r$, with $M=1$, and $\kappa=1$. Left panel for $\Lambda=0.04$, and right panel for $\Lambda=0.11$.} 
\label{Potential}
\end{figure}

\begin{figure}[h]
\begin{center}
\includegraphics[width=0.32\textwidth]{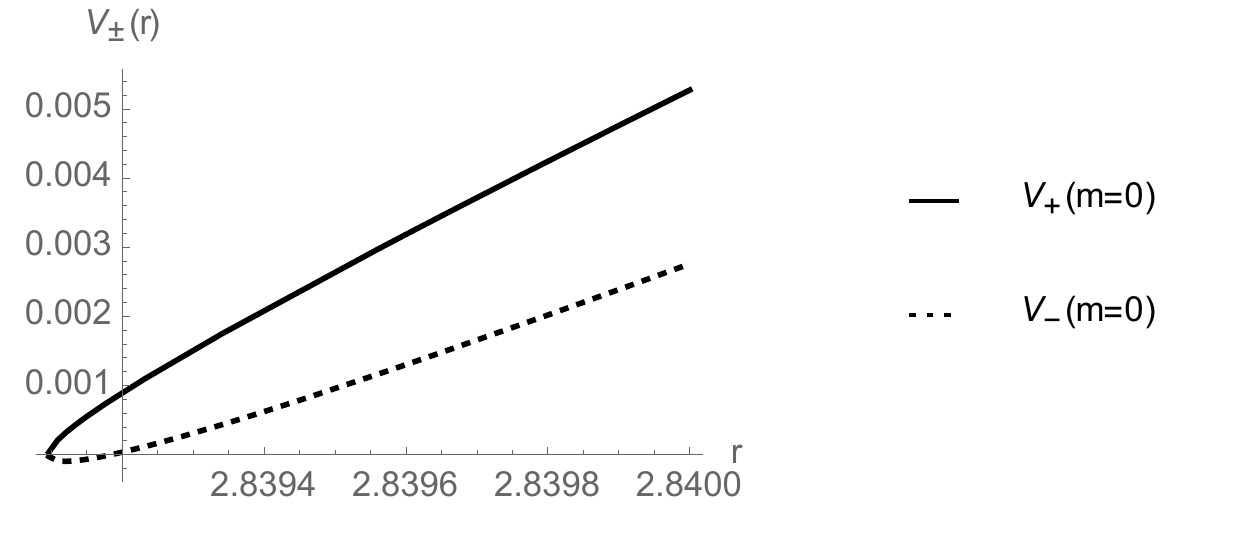}
\includegraphics[width=0.32\textwidth]{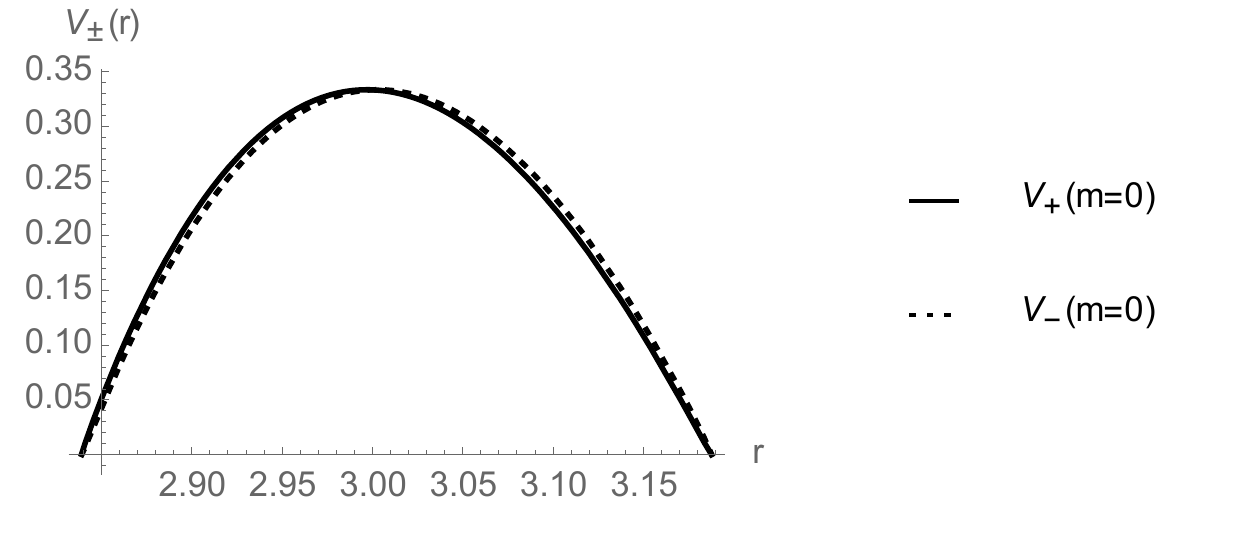}
\includegraphics[width=0.32\textwidth]{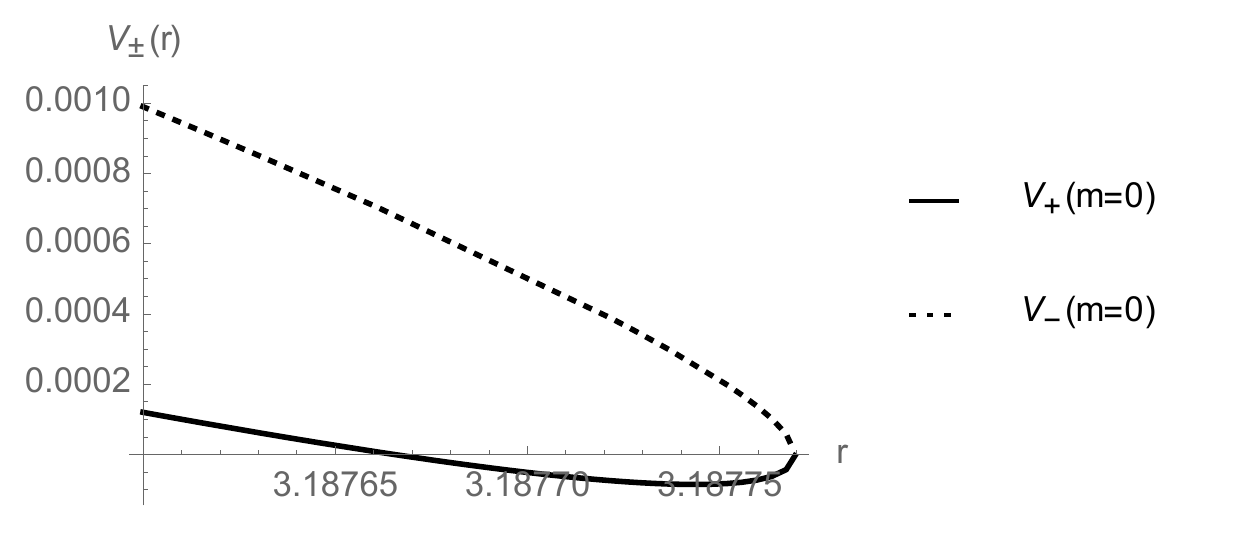}
\end{center}
\caption{The behaviour of $V_{\pm}(r)$ as a function of $r$ for massless Dirac field, with $M=1$,  $\Lambda=0.11$, and $\kappa=30$. Left panel for the region near to $r_H$, center panel for the global behaviour, and right panel for the region near to $r_{\Lambda}$.} 
\label{Potential2}
\end{figure}

The behaviour of the effective potentials as a function of $r$ is shown in Fig. \ref{Potential} for a low value of the parameter $\kappa=1$, with $M=1$, $\Lambda=0.04$ (left panel) and  $\Lambda=0.11$ (right panel), and for a higher value of $\kappa=30$ in Fig. \ref{Potential2}, with $M=1$, and $\Lambda=0.11$. The effective potentials of Fig. \ref{Potential} exhibit negative gaps, $V_-$ has a negative gap around the neighborhood of the event horizon, while that $V_+$ has a negative gap around the cosmological horizon. Usually when the effective potential is negative
in some region, growing perturbations can appear in the spectrum indicating an instability of the system under such perturbations. However in \cite{Konoplya:2020zso}  the authors studied the 
stability of massless Dirac field on Schwarzschild de Sitter black holes via time-domain integration of the scalar wave equation finding the complete stability for all $\kappa$ modes. 
On the other hand, it was also pointed out in \cite{Chang:2006kf} that the tunnelling process can occur when the energy of the Dirac field $\omega^{2}$, which is always larger than the mass $m$,
is smaller than the peak value of the effective potential $(V_{max})$,  
then the QNMs exist only when
$m^{2} < \omega^{2} < V_{max}$.

\newpage

\section{Quasinormal modes}
\label{numerical}

Now, in order to obtain the QNFs, we shall solve numerically the generalized eigenvalue equation (\ref{q1}).  We will only consider Eq. (\ref{q1}) because Eqs. (\ref{q1}) and (\ref{q2}) are isospectral, i.e., they posses the same QNM spectrum. Here, we use the pseudospectral Chebyshev method to find the QNM spectrum, see for instance \cite{Boyd}. Also, we will write only the QNFs with positive real part, because similar QNFs differing only in the sign of the real part are present in the spectrum. Now, we analyze the asymptotic behaviours of the radial function at the event and cosmological horizons along with the boundary conditions in both limits to accommodate the boundary conditions in the pseudospectral Chebyshev method. So, considering Eq. (\ref{q1}), in the vicinity of the horizon
the function $\psi_1(r)$ behaves as
\begin{equation}
\psi_1(r)=C_1 (r-r_H)^{-i \omega / f'(r_H)} +C_2 (r-r_H) ^{1/2 + i\omega /f'(r_H)} \,,
\end{equation}
where, the first term represents an ingoing wave and the second represents an outgoing wave near the black hole horizon.
So, imposing the requirement of only ingoing waves at the horizon, we fix $C_2=0$. On the other hand, at the cosmological horizon the function $\psi_1(r)$ behaves as
\begin{equation}
R(y)= D_1 (r-r_C)^{-i \omega/f'(r_C)} +D_2 (r-r_C)^{1/2+i \omega /f'(r_C)}  \,,
\end{equation}
where, the first term represents an outgoing wave and the second represents an ingoing wave near the cosmological horizon. So, imposing the requirement of only ingoing waves on the cosmological horizon requires $D_1=0$.
Therefore, taking into account the behaviour of the scalar field at the event and cosmological horizons we define the following ansatz
\begin{equation}
\psi_1 (r)= (r-r_H)^{-i \omega /f'(r_H)} (r-r_C)^{1/2+i \omega /f'(r_C)} F(r) \,.
\end{equation}
Then, by inserting the above expression for $\psi_1(r)$ in Eq. (\ref{q1}), it is possible to obtain an equation for the function $F(r)$. Also, it is convenient to perform a change of variable in order to limit the values of the radial coordinate to the range $[0,1]$, thereby we define the change of variable $y=(r-r_H)/(r_{\Lambda}-r_H)$. So, the event horizon is located at $y=0$ and the cosmological horizon at $y=1$. Thus, we have to solve an equation for a function of $y$, say $F(y)$  which is regular at both, the event and cosmological horizons. To use the pseudospectral method, $F(y)$ must be expanded in a complete basis of functions $\varphi_i(y)$: $F(y)=\sum_{i=0}^{\infty} c_i \varphi_i(y)$, where $c_i$ are the coefficients of the expansion, and we choose the Chebyshev polynomials for the complete basis of functions, which are defined by $T_j(x)= \cos (j \cos^{-1}x)$, where $j$ corresponds to the grade of the polynomial. The sum must be truncated until a $N$ value, therefore the function $F(y)$ can be approximate by  
\begin{equation}
F(y) \approx \sum_{i=0}^N c_i T_i (x)\,.
\end{equation}
Thus, the solution is assumed to be a finite linear combination of the Chebyshev polynomials,
that are well defined in the interval $[-1,1]$. Due to the variable $y$ is defined in the interval $[0,1]$, $x$ and $y$ are related by $x=2y-1$.

Then, 
the interval $[0,1]$ is discretized at the Chebyshev collocation points $y_j$ by using the so-called Gauss-Lobatto grid where
\begin{equation}
    y_j=\frac{1}{2}[1-\cos(\frac{j \pi}{N})]\,, \,\,\,\, j=0,1,...,N \,.
\end{equation}
After that, the differential equation is evaluated at each collocation point. So, a system of $N+1$ algebraic equations is obtained, which corresponds to a generalized eigenvalue problem and it can be solved numerically to obtain the QNMs spectrum, by employing the built-in Eigensystem[ ] procedure  in Wolfram’s Mathematica \cite{WM}. In this work, we use a value of $N$ into the interval [80-120] with an average running time in the range [75s-245s] which depends on the convergence of $\omega$ to the desired accuracy. We use an accuracy of eight decimal places with the exception of table \ref{tabla1}, where we use an accuracy of five decimal places. In order to guarantee the accuracy of the results the code was executed for several increasing values of $N$ until no difference was observed in the value of the QNF. Also, the complete parameter space associated to the models is $mM\geq 0$, $0<\Lambda M^2 \leq 1/9$, and $\kappa= \pm (\ell+1)$, with $\ell=0,1,2, ...$. Here, the regions of the parameter space explored is $0 \leq mM \leq 0.20$ due to the anomalous decay rate behaviour is observed for small values of $mM$, $0<\Lambda M^2\leq 0.11$, and a discrete set of values of $\kappa$ in the interval [1,200].
\newline

\begin{figure}[h]
\begin{center}
\includegraphics[width=0.6\textwidth]{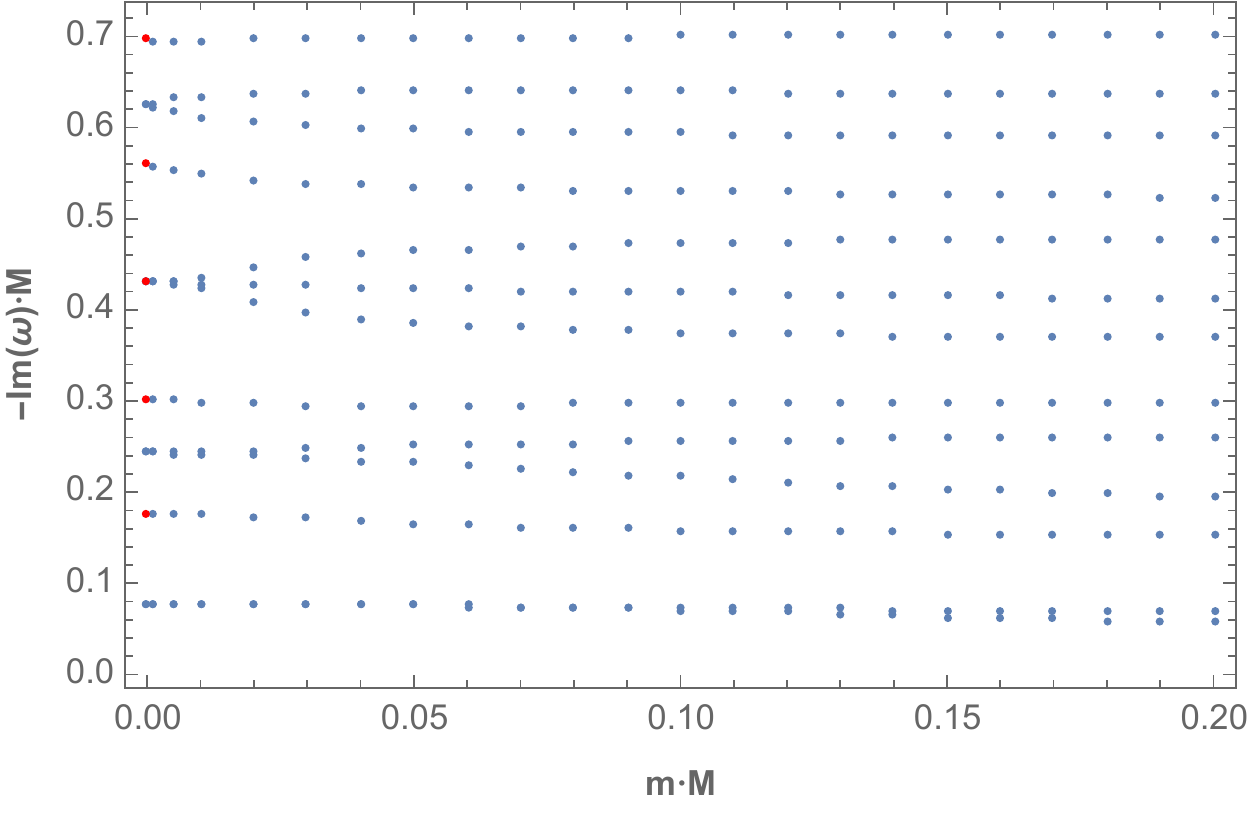}\\
\includegraphics[width=0.6\textwidth]{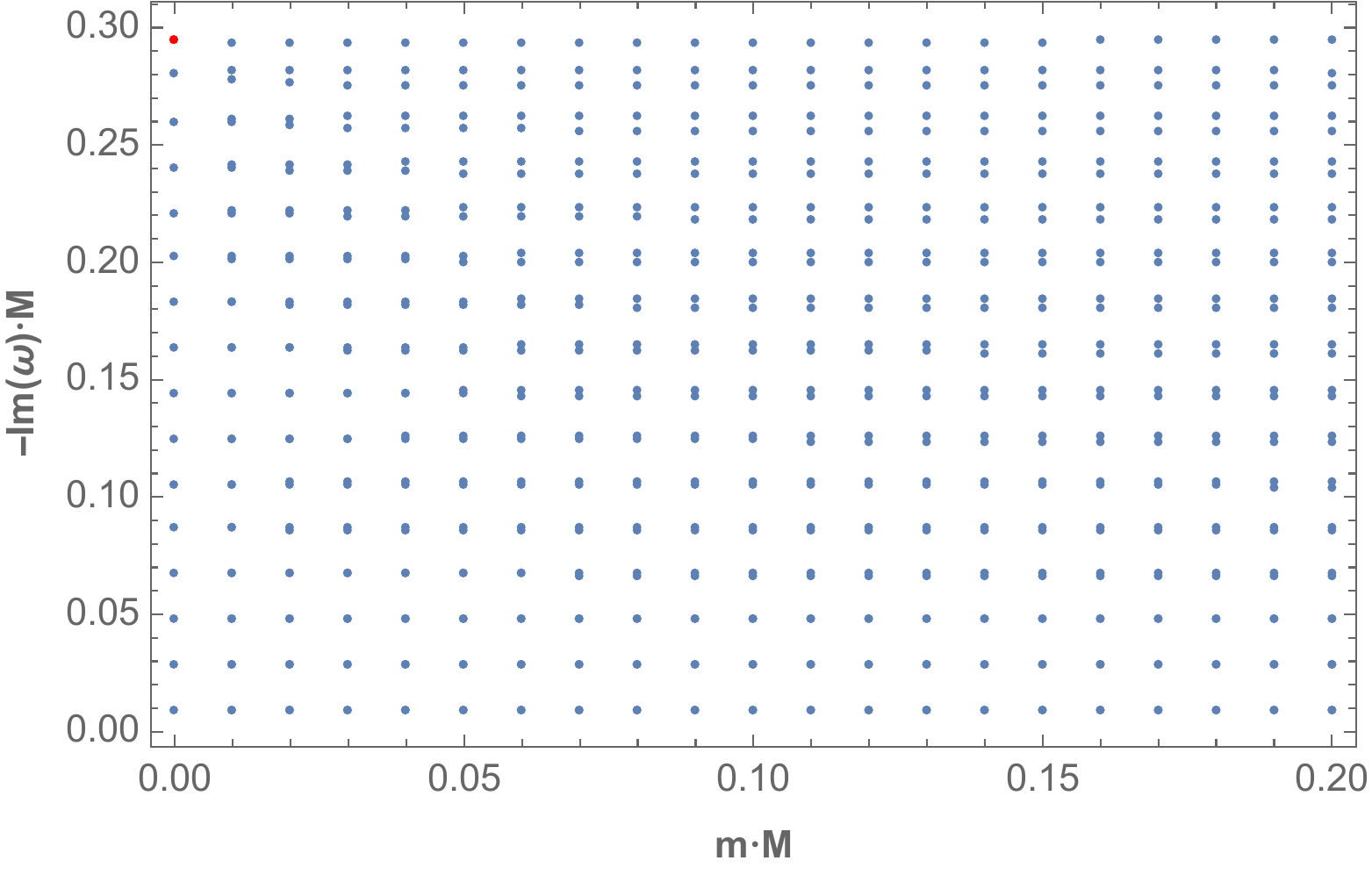}
\end{center}
\caption{The behaviour of $-Im(\omega) M$ as a function of $m M$, {\bf{for $\psi_1(r)$}} with $\kappa=1$. Top panel for $\Lambda M^2=0.04$, and bottom panel for $\Lambda M^2=0.11$. The red points correspond to purely imaginary QNFs while that blue points correspond to complex QNFs.} 
\label{FMCA}
\end{figure}

Now, in order to identify the different families of modes that are present in the fermionic spectrum, we plot in Fig. \ref{FMCA} the behaviour of $-Im(\omega) M$ versus the product of the fermionic field mass and black hole mass $m M$. The red points correspond to purely imaginary QNFs while that blue points correspond to complex QNFs. We can note, the existence of two family of modes. One of them, correspond to the de Sitter (dS) modes ($\omega_{dS}$), in that they continuously approach those of empty de Sitter space in the limit that the black hole mass vanishes. In this case, the QNFs for massless fermionic field are purely imaginary and they acquire a real part when the fermionic field is massive. The other family corresponds to the photon sphere (PS) modes ($\omega_{PS}$), that are complex frequencies, and continuously approach those of Schwarszchild black hole when $\Lambda \rightarrow 0$. In the following we will study both families separately. Also, note that in Fig. \ref{FMCA} the fine structure related to the coupling between the chirality and the mass of the field appears spontaneously in the spectrum. 

\newpage

\subsection{Photon sphere modes}

\subsubsection{Massless fermionic fields}

\begin{figure}[h]
\begin{center}
\includegraphics[width=0.46\textwidth]{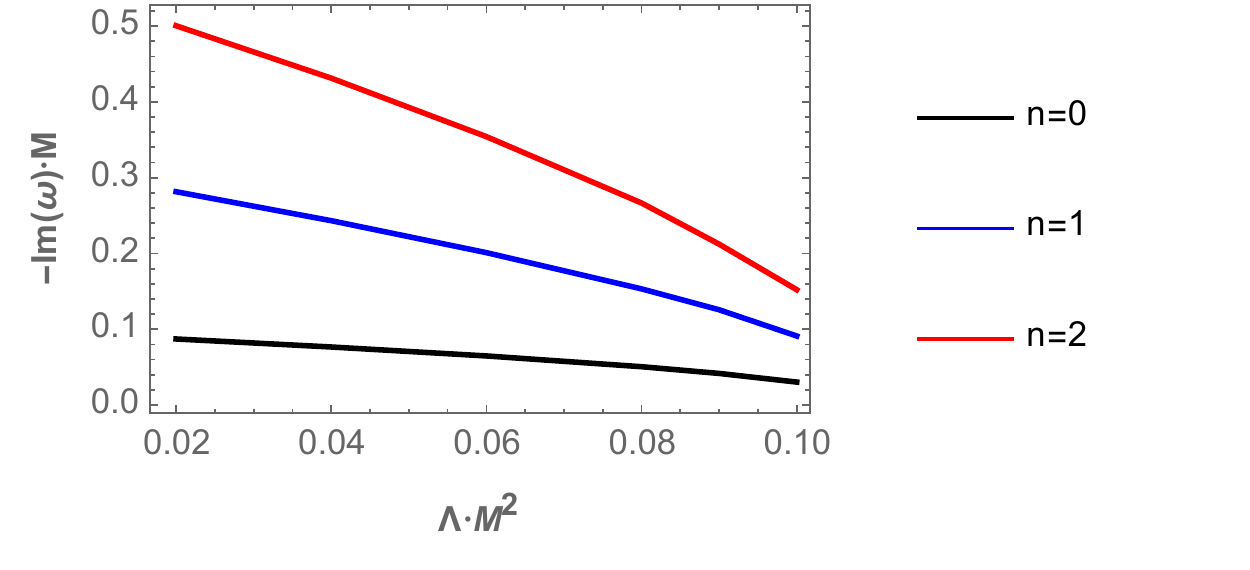}
\includegraphics[width=0.46\textwidth]{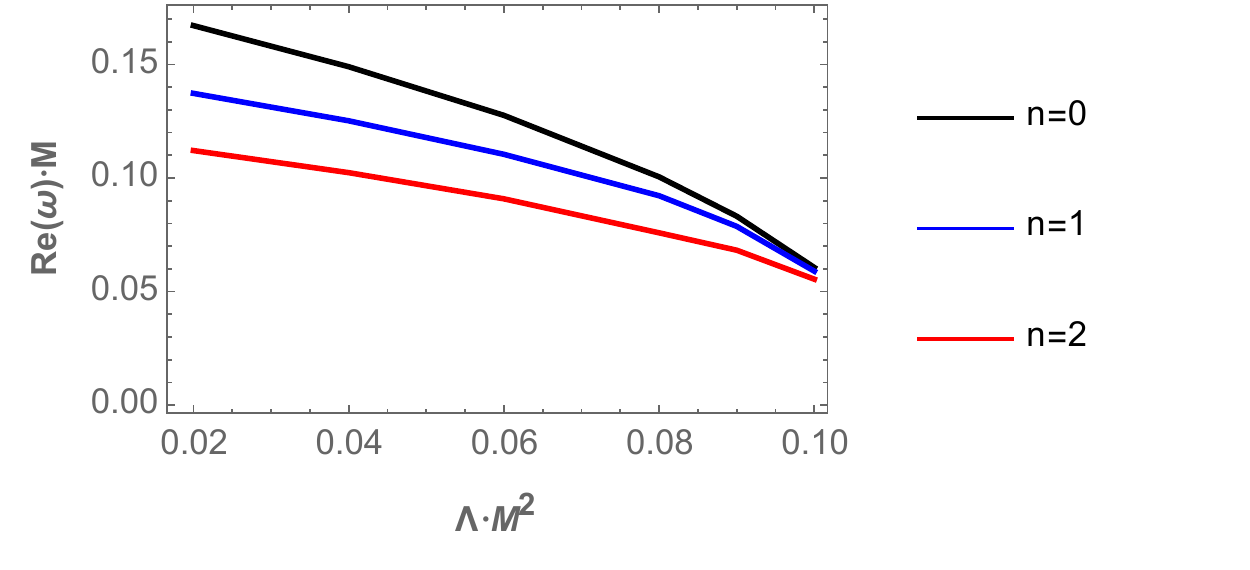}
\end{center}
\caption{The behaviour of $-Im(\omega_{PS}) M$ (left) and $Re(\omega_{PS}) M$ (right) as a function of $\Lambda M^2$ for massless fermionic fields with $\kappa=1$, and different values of the overtone number $n_{PS}=0,1,$ and $2$.} 
\label{FMC}
\end{figure}

Now, we plot in Fig.\ref{FMC}, the behaviour of $-Im(\omega_{PS}) M$ (left panel) and  $Re(\omega_{PS}) M$ (right panel) as a function of $\Lambda M^2$, for massless fermionic fields and for the overtone numbers $n_{PS}=0,1,$ and $2$. We can observe that when the cosmological constant increases the decay rate and the frequency of the oscillation decreases. However, when the overtone number increases the decay rate increases and the frequency of the oscillations decreases.  

On the other hand, it is known that the photon sphere modes are given by
\begin{equation}
\label{QN}
\omega_{PS}=\frac{\sqrt{1-9M^2\Lambda}}{3\sqrt{3}M}\left(\ell-(n_{PS}+\frac{1}{2})i\right) \,,
\end{equation}
which was 
obtained using the 1st-order WKB method \cite{Zhidenko:2003wq}. Thus, in order to check the correctness and accuracy of the numerical technique with respect to previous results Eq. (\ref{QN}), we show in Table \ref{QNMC}, the values obtained via the pseudospectral Chebyshev method and using Eq. (\ref{QN}) for high values of $\kappa$. Also, we show the relative error, which is defined by
\begin{equation}
\label{ERe}
\epsilon_{Re(\omega)} =\frac{\mid Re(\omega_1)- Re(\omega_0)\mid}{Re(\omega_0)}\cdot 100\%
\end{equation}
\begin{equation}
\label{EIm}
\epsilon_{Im(\omega)} =\frac{\mid Im(\omega_1)- Im(\omega_0)\mid}{Im(\omega_0)} \cdot 100\%
\end{equation}
where $\omega_1$ corresponds to the result from Eq.  (\ref{QN}) \cite{Zhidenko:2003wq}, and $\omega_0$ denotes our result with the pseudospectral Chebyshev method.  We can observe that the error does not exceed $0.006$ $\%$ in the imaginary part, and $0.023$ $\%$ in the real part. Also, as it was observed, the frequencies all have a negative imaginary part, which means that the propagation of massless fermionic field is stable in this background. 

\begin {table}[ht] 
\caption {Quasinormal frequencies $\omega_{PS} M$ for massless fermionic fields with $\kappa = 50,100,125,150,175$, and $200$ in the background of Schwarzschild dS black holes, with $\Lambda M^2 = 0.04 $.}
\centering
\scalebox{0.85}{
\begin {tabular} { | c | c | c | c |c |}
\hline
\multicolumn{5}{ |c| }{$n_{PS}=0$} \\ \hline
$\kappa$ &  \text{ Pseudospectral method}  & $ WKB $ & $ \epsilon_{Re(\omega)} $ & $ \epsilon_{Im(\omega)} $\\\hline
$50$ &
$7.69790323 - 0.07697990 i$ & $7.69800359 - 0.07698004 i$ & $0.00130373$
& $0.00018187 $\\
$100$ &
$15.39595700 - 0.07698000 i$ & $15.39600718-0.07698004i$ & $0.00032593$ &
$0.00005196 $\\
$125$ &
$19.24496883 - 0.07698001 i$ & $19.24500897-0.07698004 i$ &$0.00020857$ &
$0.00003897$\\
$150$ &
$23.09397731 - 0.07698002 i$ & $23.09401077-0.07698004i$ &
$0.00014489$ &
$0.00002598$\\
$175$ &
$26.94298389 - 0.07698002 i$ & $26.94301256-0.07698004i$&
$0.00010641$ &
$0.00002598$\\
$200$ &
$30.79198927 - 0.07698003 i$ & $30.79201436-0.07698004i$&
$0.00008148$ &
$0.00001299 $ \\
\hline
\multicolumn{5}{ |c| }{$n_{PS}=1$} \\ \hline
$\kappa$ & \text{Pseudospectral method} & $ WKB $ & $ \epsilon_{Re(\omega)} $ & $ \epsilon_{Im(\omega)} $ \\\hline
$50$ &
$7.69735582-0.23094427i$ & $7.69800359-0.23094011i$ &
$0.00841549$ &
$0.00180130 $ \\
$100$ &
$15.39568329 - 0.23094115 i$ & $15.39600718-0.23094011 i$&
$0.00210377$ &
$0.00045033 $ \\
$125$ &
$19.24474986 - 0.23094077 i$ & $19.24500897-0.23094011 i$&
$0.00134639$ &
$0.00028579$\\
$150$ &
$23.09379484 - 0.23094057 i$ & $23.09401077-0.23094011 i$&
$0.00093501 $ &
$0.00019919 $ \\
$175$ &
$26.94282748 - 0.23094045 i$ & $26.94301256-0.23094011 i$&
$0.00068694$ &
$0.00014722$  \\
$200$ &
$30.79185241 - 0.23094037 i$ & $30.79201436-0.23094011 i$&
$0.00052595$&
$ 0.00011258 $ \\
\hline
\multicolumn{5}{ |c| }{$n_{PS}=2$} \\ \hline
$\kappa$ & \text{Pseudospectral method} & $ WKB $ & $ \epsilon_{Re(\omega)} $ & $ \epsilon_{Im(\omega)} $ \\\hline
$50$ &
$7.69626106 - 0.38492238 i$ & $7.69800359-0.38490011 i $&
$0.02264125$ &
$0.00578558$ \\
$100$ &
$15.39513589 - 0.38490573 i$ &  $15.39600718-0.38490011 i $&
$0.00565952$ &
$0.00146010$ \\
$125$ &
$19.24431194 - 0.38490373 i$ &  $19.24500897-0.38490011 i $&
$0.00362201$ &
$0.00094049$ \\
$150$ &
$23.09342991 - 0.38490265 i$ &  $23.09401077-0.38490011 i $&
$0.00251526 $ &
$0.00065991 $\\
$175$ &
$26.94251468 - 0.38490199 i$ &  $26.94301256-0.38490011 i $&
$0.00184793 $ &
$0.00048844 $ \\
$200$ &
$30.79157871 - 0.38490157 i$ &  $30.79201436-0.38490011 i $ &
$0.00141484$ &
$0.00037932$ \\
\hline
\end {tabular}}
\label{QNMC}
\end {table}

\newpage

\subsubsection{Massive fermionic fields}

In this section, we will consider the propagation of massive fermionic fields, as we will see, the two chiralities of massive fermions lead to an additional fine structure in the spectrum, as the one reported for Schwarzschild and Kerr backgrounds by using the convergent Frobenius method \cite{Konoplya:2017tvu}, and contrary to massless case, where the fine structure is not present, thereby the coupling between the chirality and the mass of the field leads to a fine structure in the spectrum of the QNMs. Also, we will show that the anomalous behaviour of the decay rate of the fermionic QNMs is present in this family, depending on the values of $\Lambda M^2$, as well as a critical fermionic field mass.  \\

\begin{figure}[h]
\begin{center}
\includegraphics[width=0.4\textwidth]{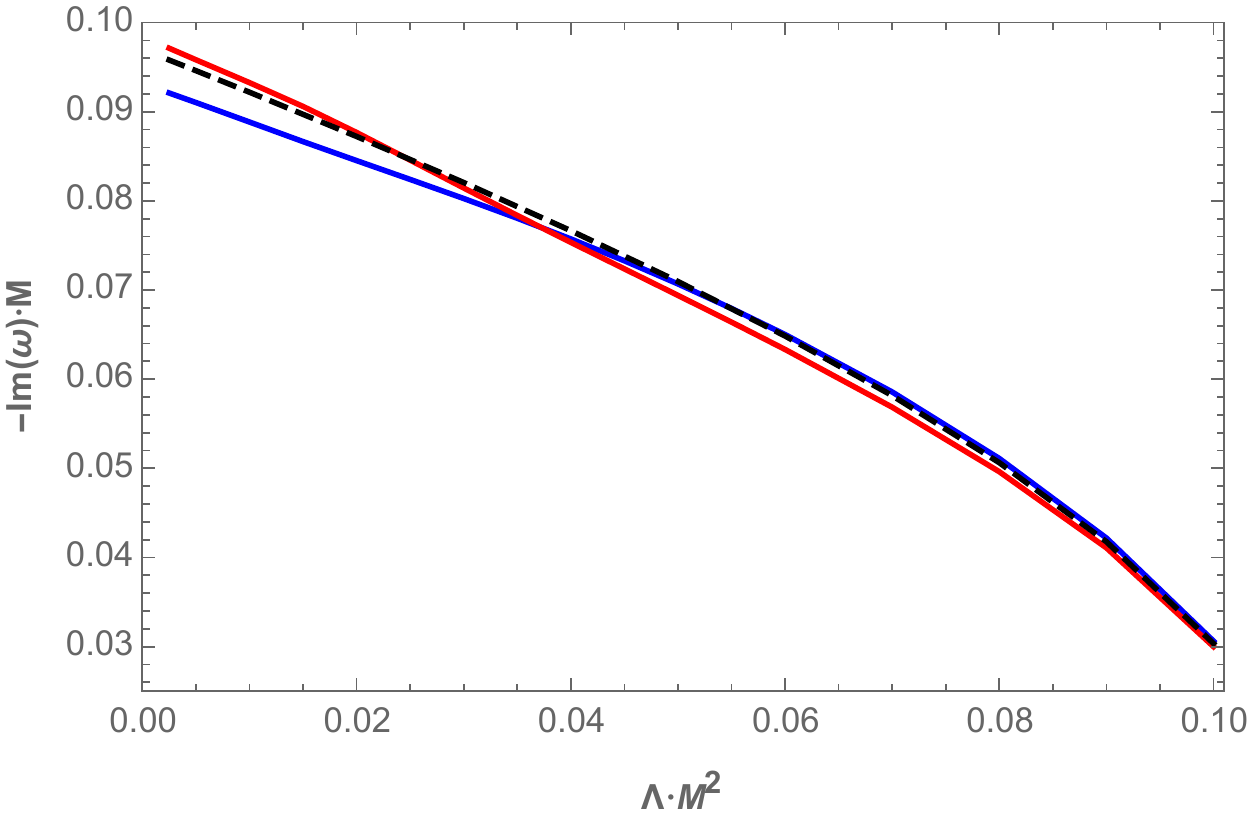}
\includegraphics[width=0.4\textwidth]{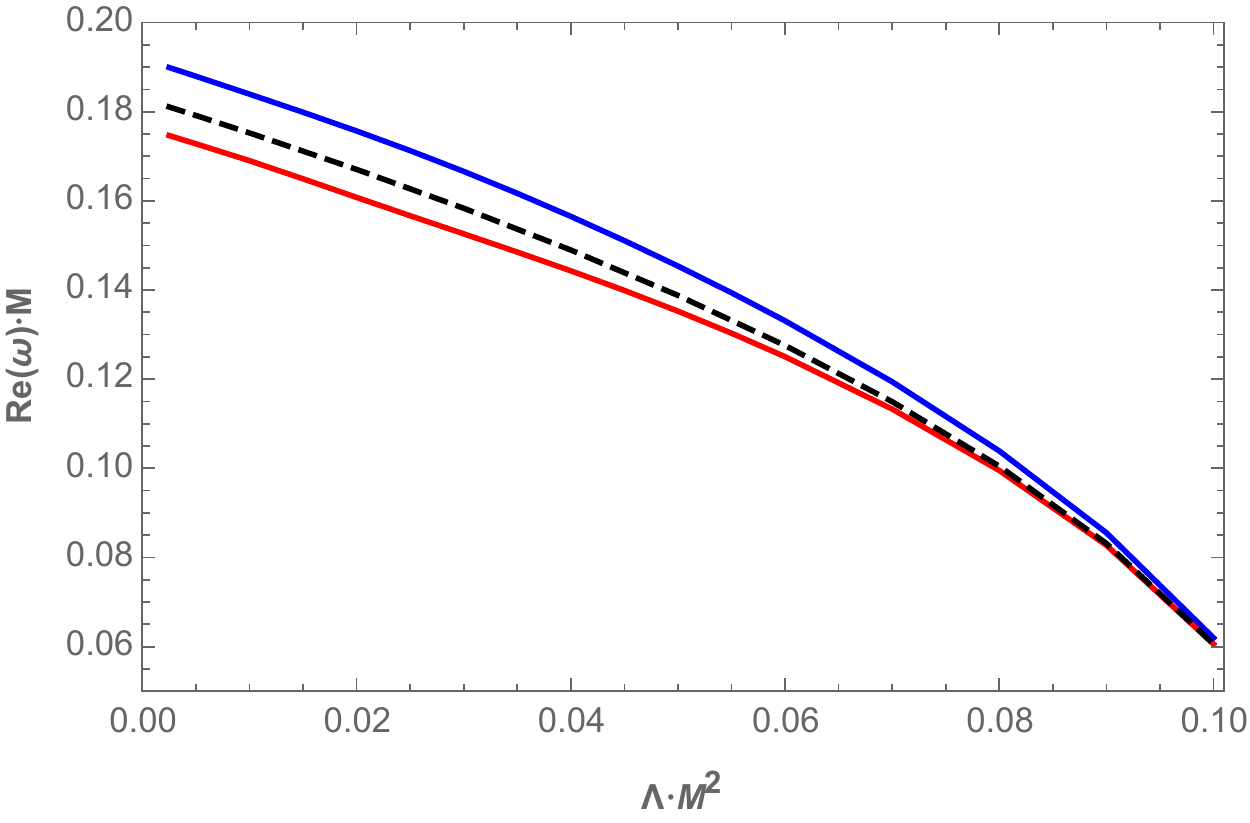}
\end{center}
\caption{The behaviour of $-Im(\omega_{PS})M$ (left panel), and $Re(\omega_{PS})M$ (right panel) obtained through the pseudospectral Chevyshev method, as a function of $\Lambda M^2$, for $\kappa=1$. Black dashed line for $mM=0$, blue line for $mM=0.05$ and positive chirality, and red line for $mM=0.05$ and negative chirality.}
\label{Q}
\end{figure}

{\bf{Fine structure}}. As we mentioned, it is possible to observe a spontaneous split of the QNFs by using the pseudospectral Chevyshev method when $m \neq 0$, which leads a fine structure in the fermionic spectrum  and it is associated to the two chiralities of the field. 
In spite of the specific representation used in the Dirac equation does not allow to appreciate the positive and negative chirality easily. We consider, the modes with
positive chirality as those with higher oscillation frequency, which was shown for $\Lambda = 0$ in Ref. \cite{Konoplya:2017tvu}, and it allows us to distinguish between both chiralities. In Fig. \ref{Q} we plot the behaviour of $-Im(\omega_{PS})M$ (left panel) and $Re(\omega_{PS})M$ (right panel) as a function of $\Lambda M^2$ for $n_{PS}=0$ and $m M=0, 0.05 $, the numerical values are given in Table \ref{tabla1}. So, by considering as fine structure the splitting 
generated in the QNFs, which occurs for $m M \neq 0$, in Fig.  \ref{Q} we show the effect of $\Lambda M^2$ on the fine structure for $mM =0.05$, where we can observe that the splitting
in the real part (right panel), decreases when $\Lambda M^2$ increases, and it go to zero when the black hole becomes extremal. 
For the imaginary part(left panel), we observe that there is a value of $\Lambda M^2$ where the curves intersect, i.e the fine structure vanishes. Before this point, the fine structure decreases when $\Lambda M$ increases and after it the behaviour is inverted. Then, the fine structure decreases and vanishes again when the black hole becomes extremal. For $mM=0$ we do not observe a fine structure (black curve).

\begin {table}[ht]
\caption {Quasinormal frequencies $\omega_{PS} M$ obtained through the pseudospectral Chevyshev method for  fermionic fields with $\kappa = 1$,  in the background of Schwarzschild-dS black holes.}
\centering
\begin {tabular} { | c | c | c |} \hline
$n_{PS}=0$ &  $m M=0$  & $m M=0.05$  \\\hline
$\Lambda M^2$ &
  & positive chirality     {       } \,    negative chirality \\ \hline
$0.0025$ &
$0.18107-0.09580 i$ & $0.18992-0.09210i$ \text{   } $0.17466-0.09712i $ \\ \hline
$0.005$ &
$0.17916 - 0.09461 i$ & $0.18796 -0.09103 i$ \text{   } $0.17280-0.09583 i$ \\ \hline
$0.01$ &
$0.17524 -0.09219i$ & $0.18395-0.08885i$ \text{   } $0.16901-0.09326i$ \\ \hline
$0.02$ &
$ 0.16706-0.08723i  $ & $0.17568-0.08452i$ \text{   } $0.16078-0.08767i$ \\ \hline
$0.03$ &
$0.15835-0.08206i$ & $0.16661-0.08027i$ \text{   } $0.15261-0.08145i$ \\ \hline
$0.04$ &
$0.14898-0.07666i$ & $0.15652-0.07573i$ \text{   } $0.14431-0.07535i$  \\ \hline
$0.05$ &
$0.13879-0.07096i$ & $0.14538-0.07069i$  \text{   }     $0.13526-0.06940i$ \\ \hline
$0.06$ &
$0.12756-0.06485i$ & $0.13310-0.06501i$ \text{   } $0.12504-0.06334i$ \\ \hline
$0.07$ &
$0.11496-0.05816i$ & $0.11942-0.05857i$ \text{   } $0.11330-0.05686i$ \\ \hline
$0.08$ &
$0.10048-0.05064i$ & $0.10389-0.05111i$  \text{   }  $0.09952-0.04962i$ \\ \hline
$0.09$ &
$0.08315-0.04177i$ & $0.08554-0.04218i$ \text{   } $0.08274-0.04108i$ \\ \hline
$0.1$ &
$0.06058-0.03036i$ & $0.06198-0.03062i$ \text{   }  $0.06059-0.03000i$ \\ \hline
\end {tabular} \label{tabla1}
\end {table}

\newpage    

\begin{figure}[h]
\begin{center}
\includegraphics[width=0.4\textwidth]{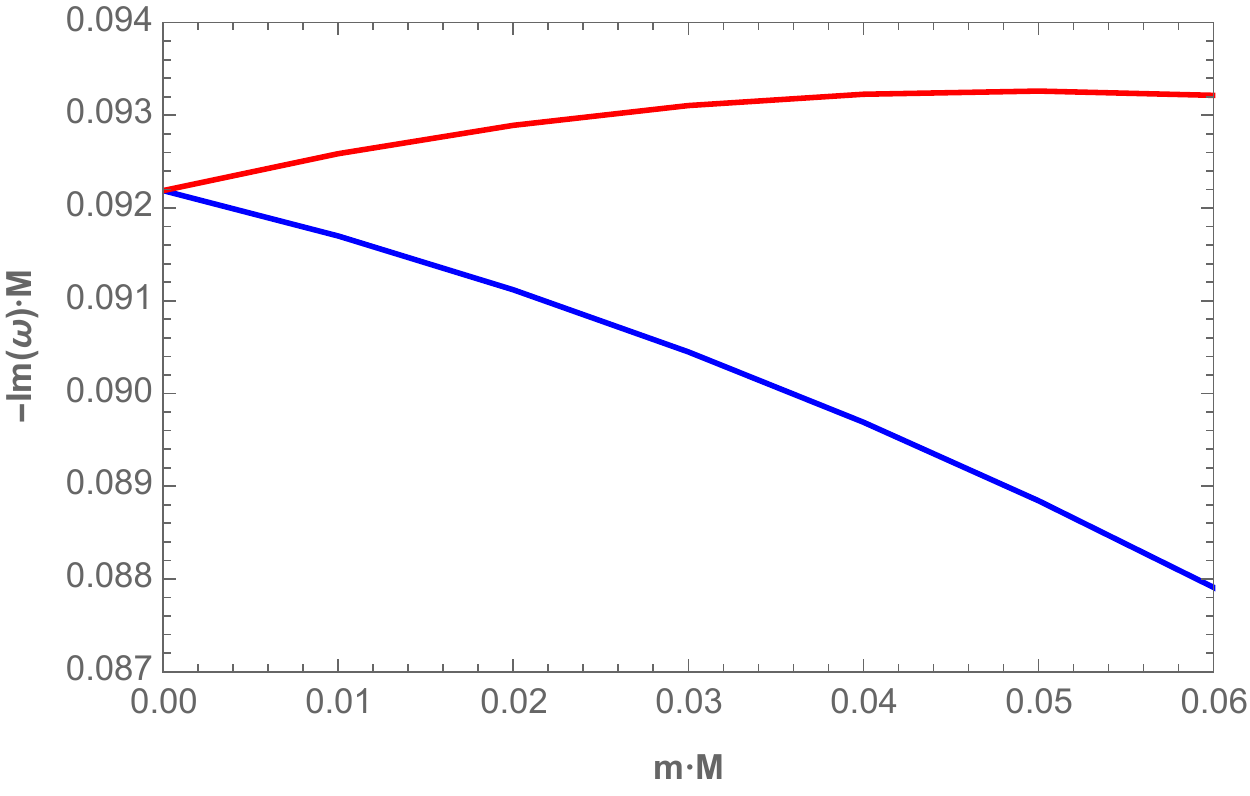}
\includegraphics[width=0.4\textwidth]{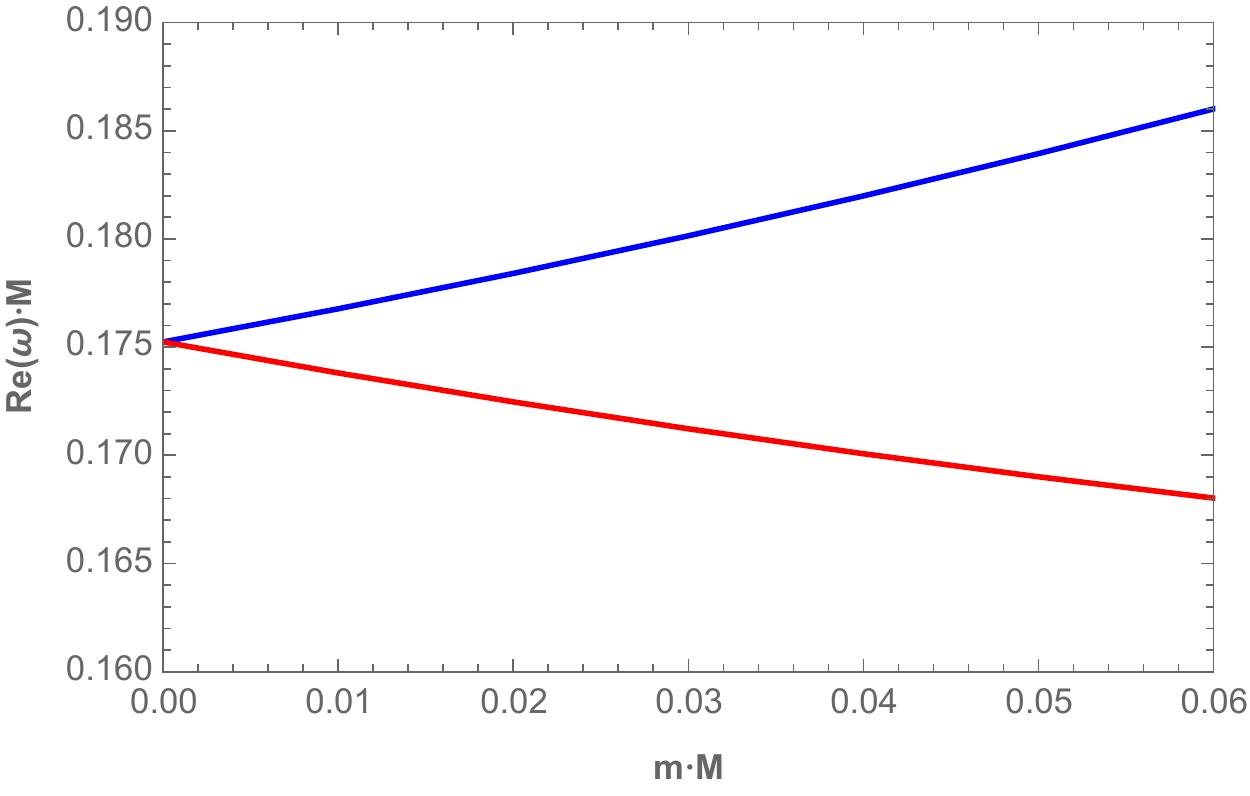}
\includegraphics[width=0.4\textwidth]{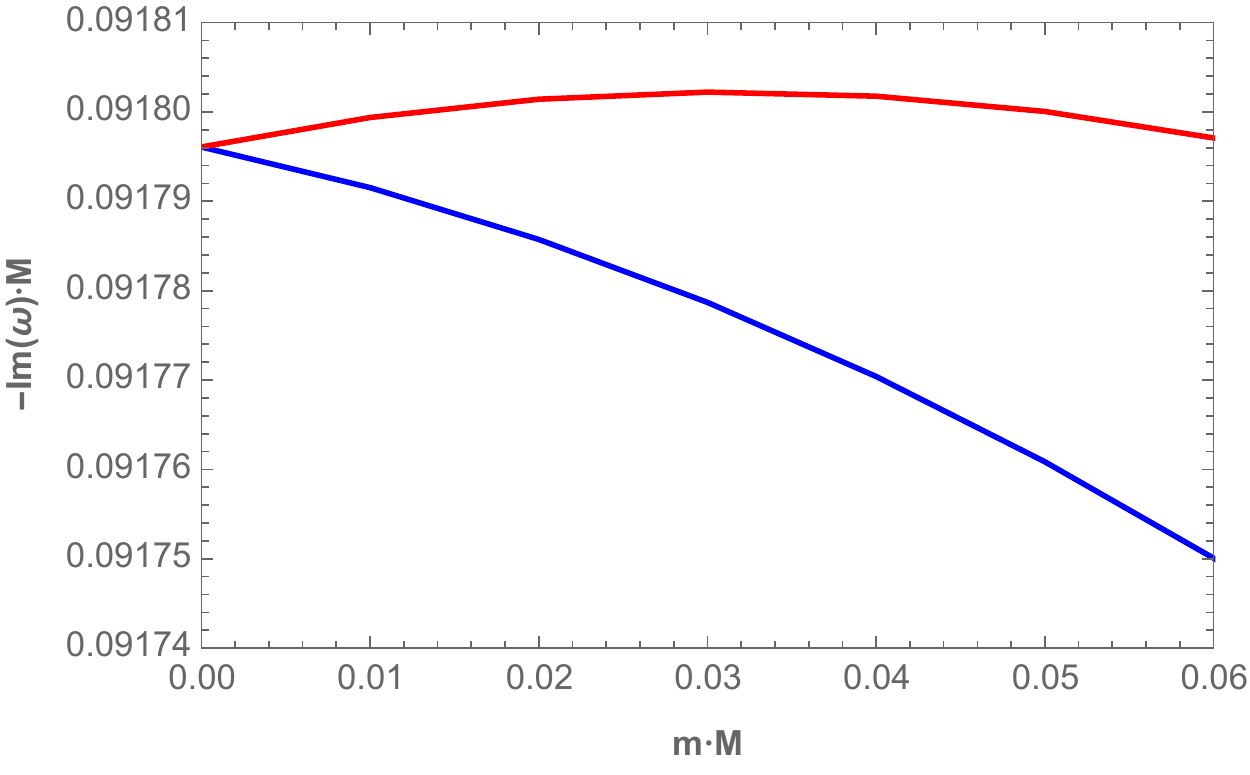}
\includegraphics[width=0.4\textwidth]{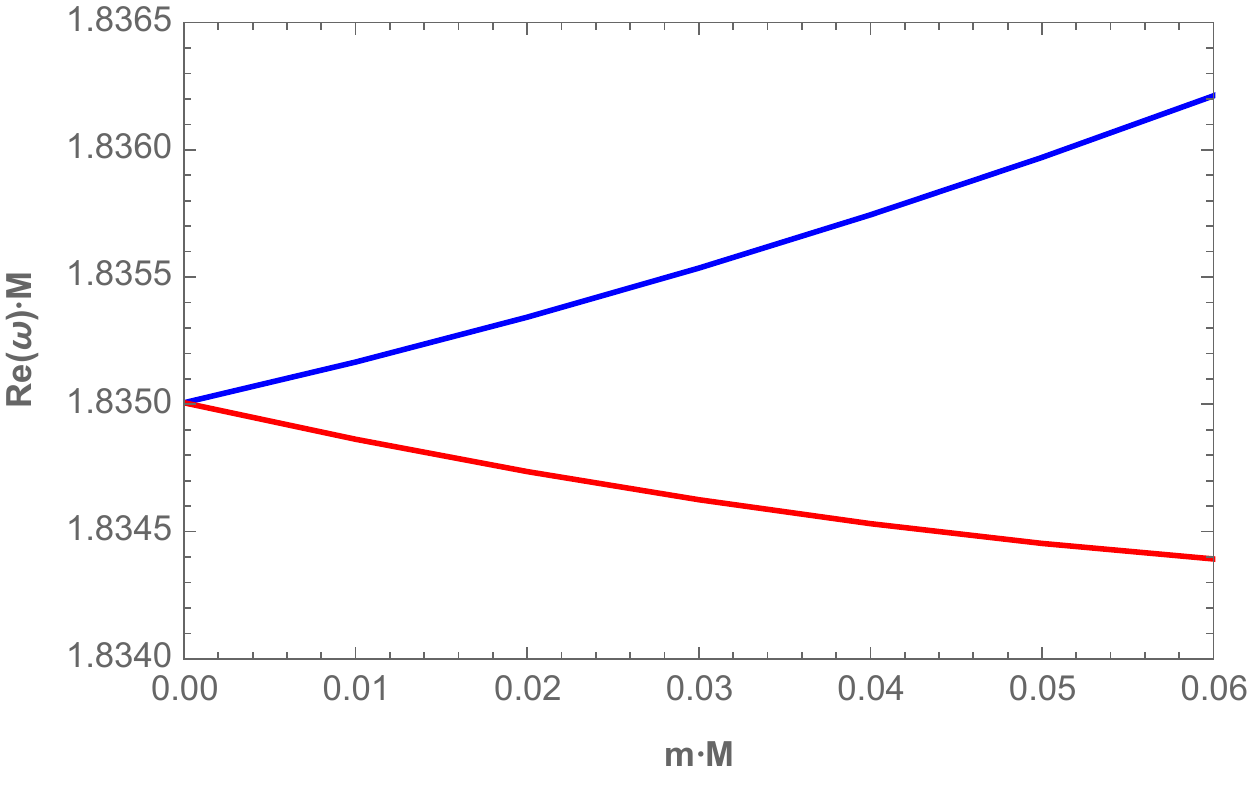}
\end{center}
\caption{The behaviour of $-Im(\omega_{PS})M$ (left panels), and $Re(\omega_{PS})M$ (right panels) obtained through the pseudospectral Chevyshev method, as a function of $mM$, for $\kappa=1$ (top panels), and $\kappa=10$ (bottom panels), with $\Lambda M^2=0.01$. Blue line for positive chirality and red line for negative chirality.}%
\label{FS}
\end{figure}

Now, in order to study the effect of $mM$ on the fine structure, we plot in Fig. \ref{FS} the behaviour of the QNFs versus $mM$ for $\Lambda M^2=0.01$ and $\kappa=1$ (top panels), and $\kappa=10$ (bottom panels), for small values of $mM$. So,
we can observe that the fine structure, in the imaginary part, decreases when the parameter $\kappa$ increases, or in other words, the fine structure is finer for higher values of $\kappa$. However, it increases with $mM$, and the same behaviour occurs  for the real part of the fine structure.

\newpage

\begin{figure}[h]
\begin{center}
\includegraphics[width=0.4\textwidth]{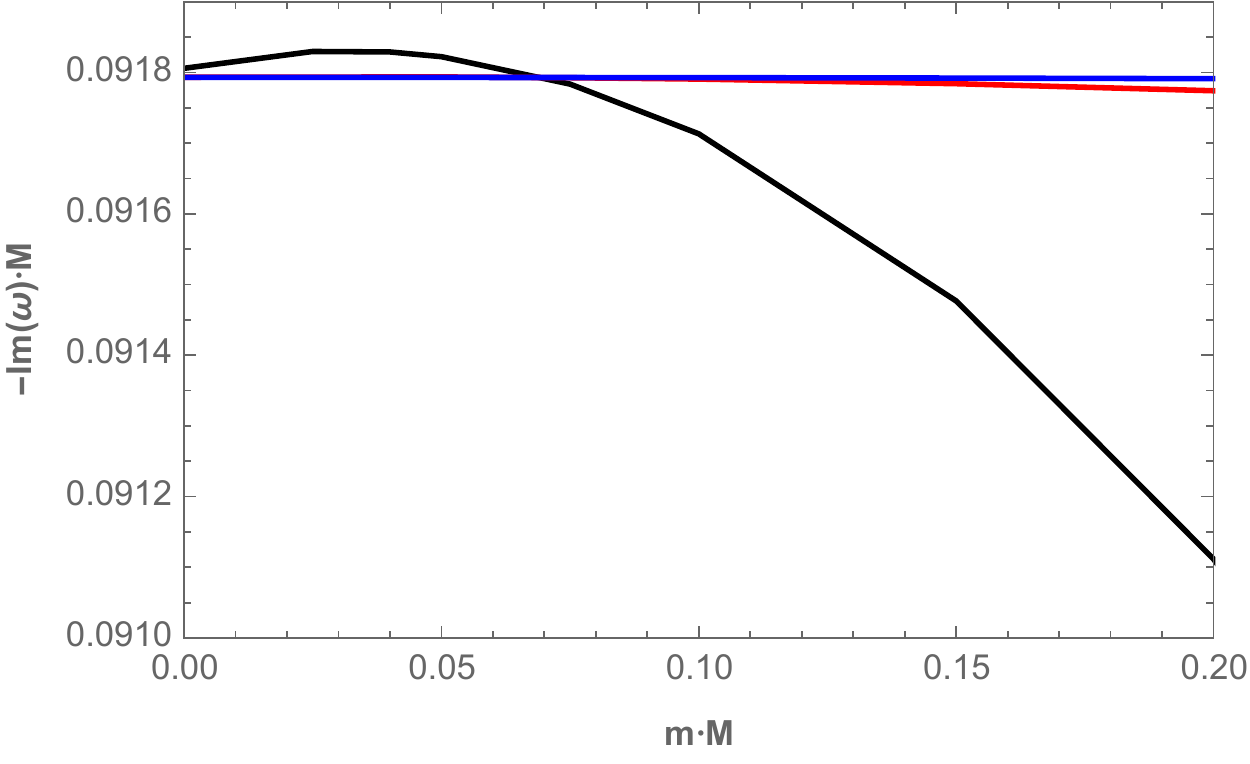}
\includegraphics[width=0.4\textwidth]{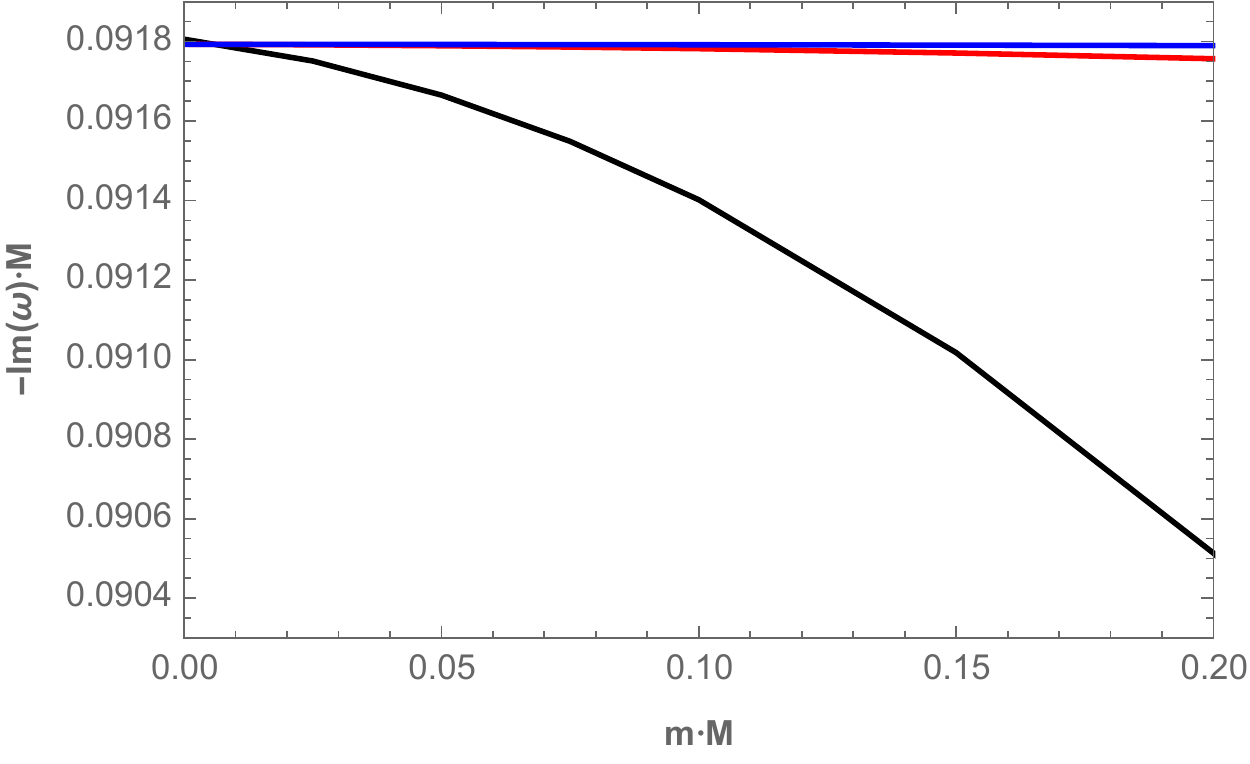}\\
\includegraphics[width=0.4\textwidth]{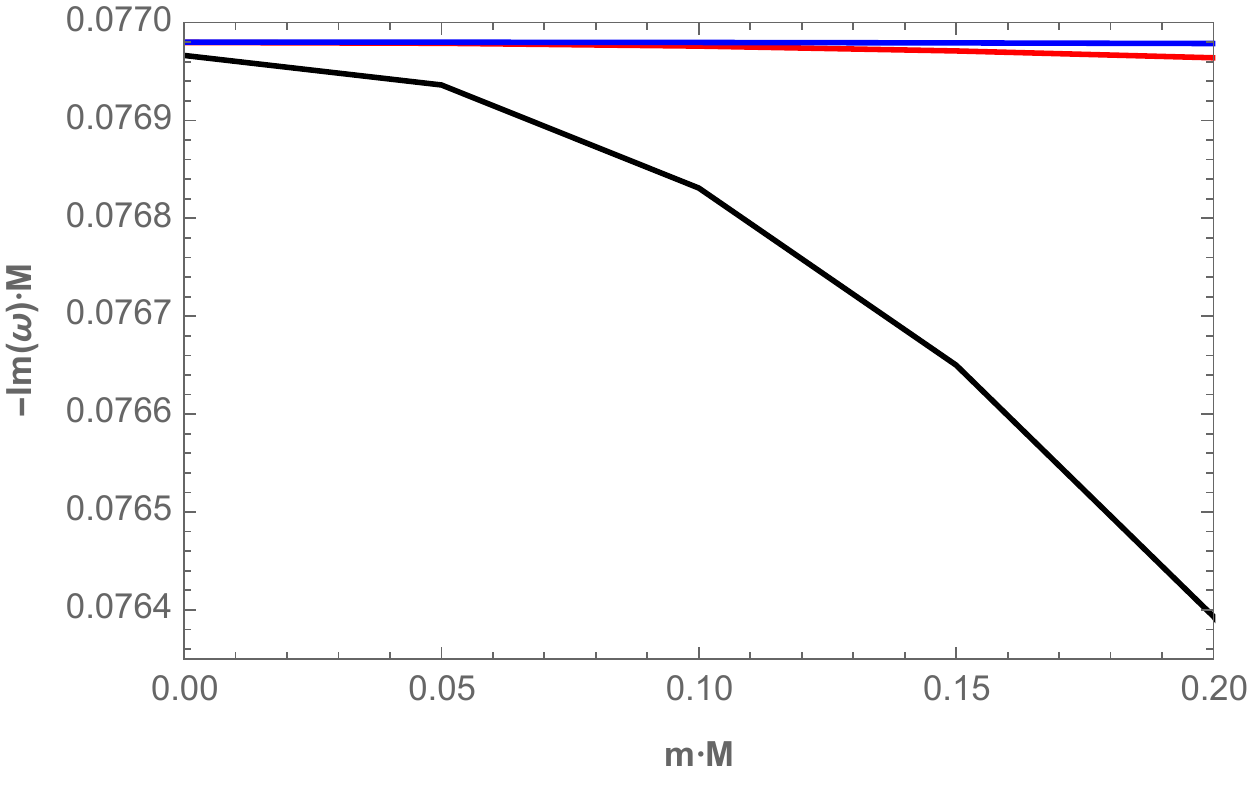}
\includegraphics[width=0.4\textwidth]{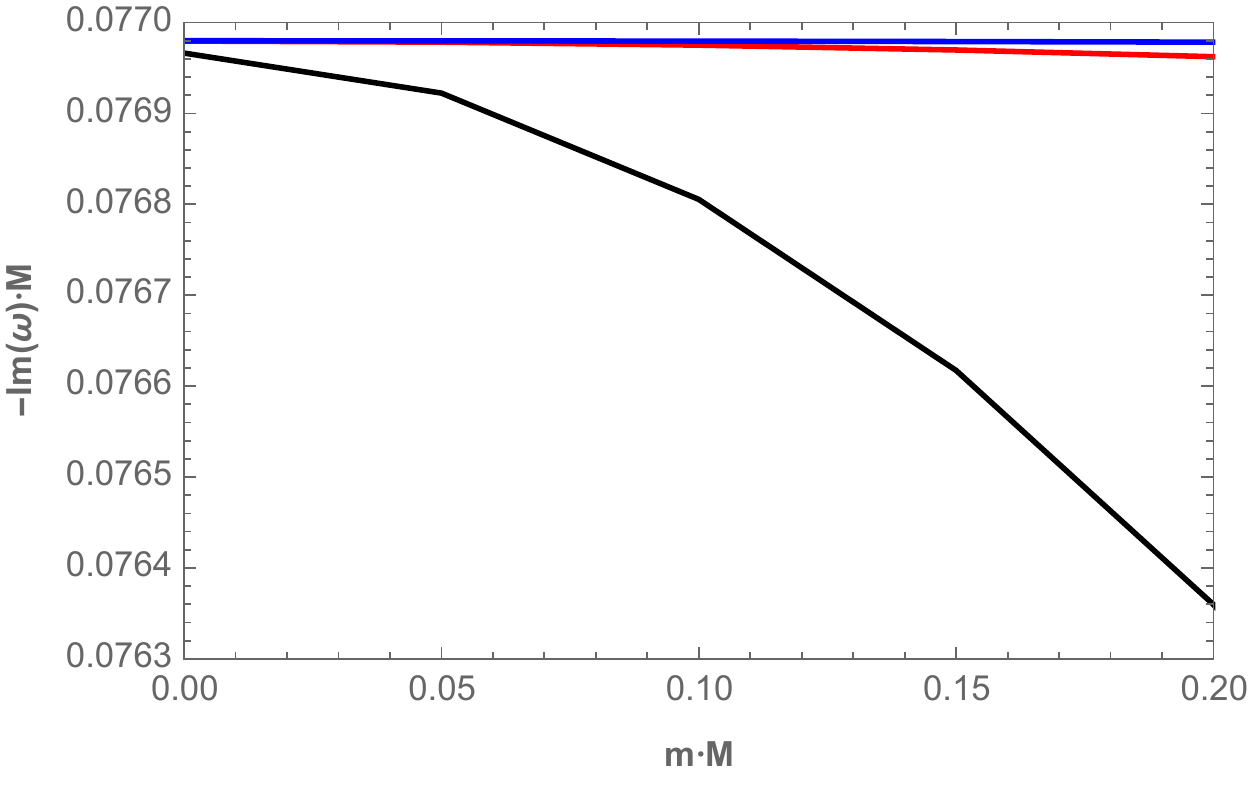}\\
\includegraphics[width=0.4\textwidth]{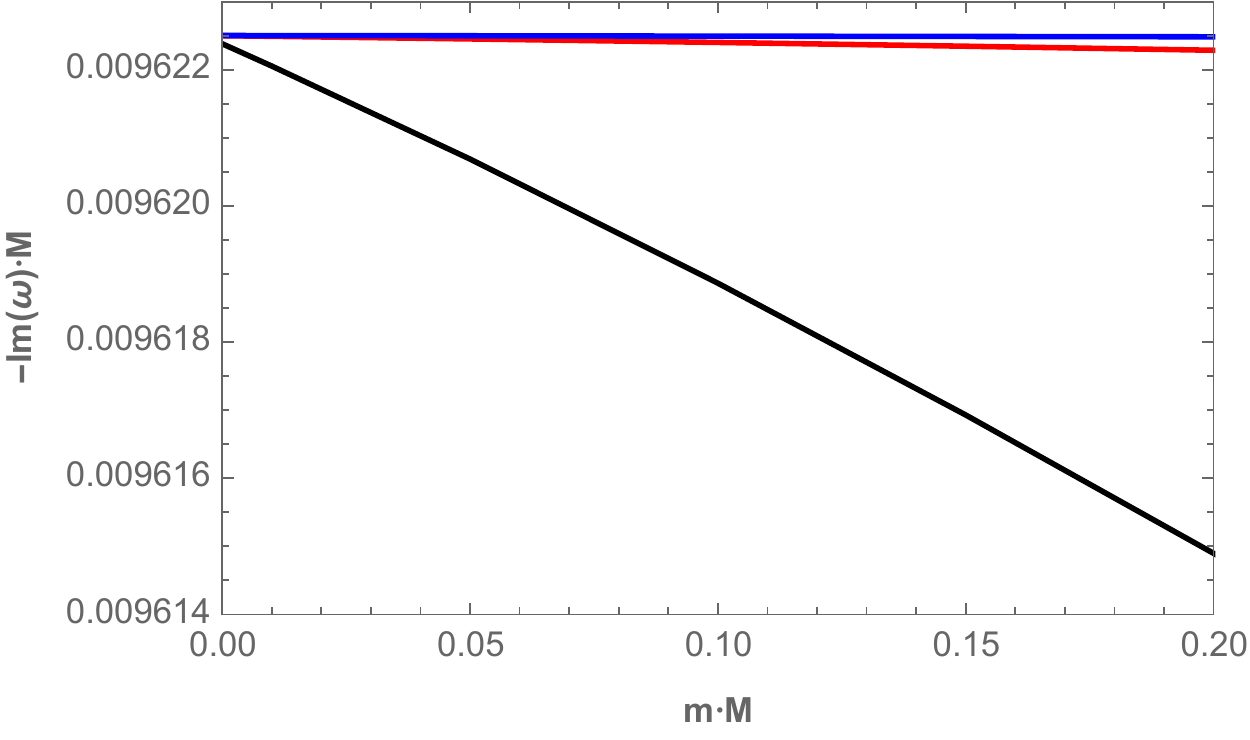}
\includegraphics[width=0.4\textwidth]{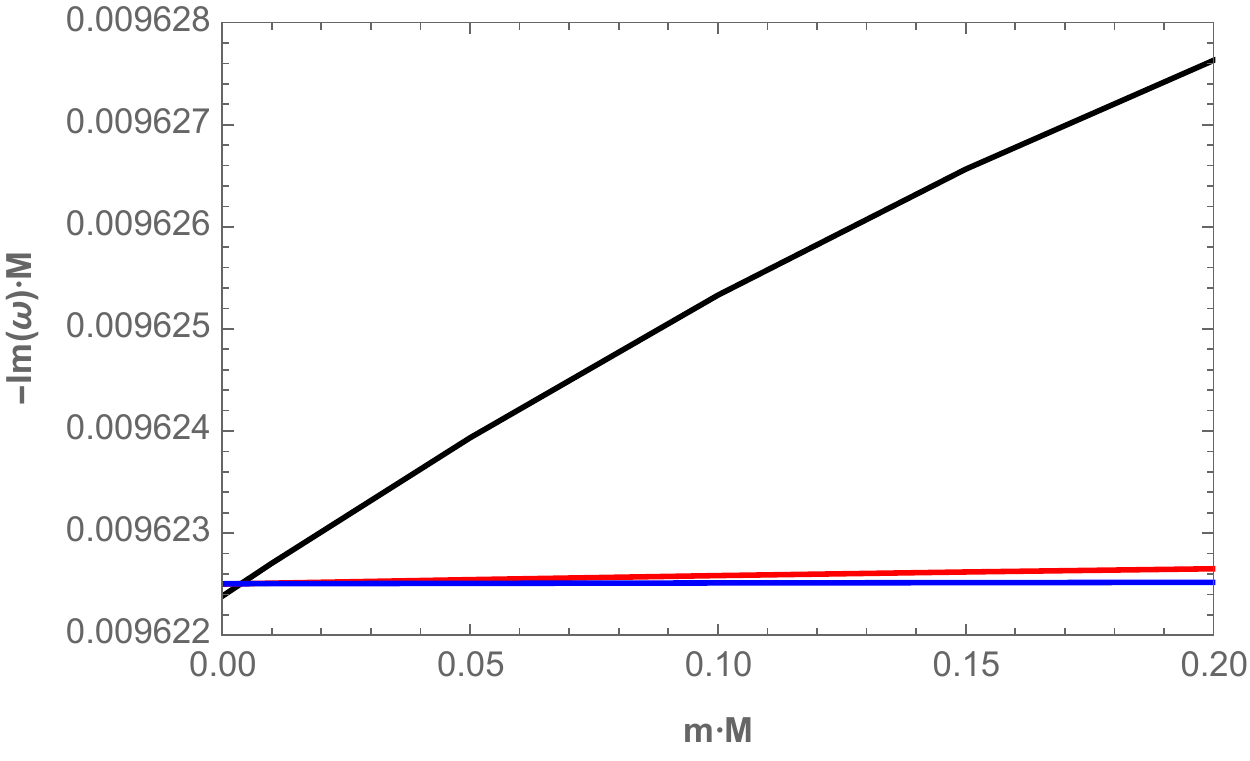}
\\
\end{center}
\caption{The behaviour of $-Im(\omega_{PS}) M$ obtained through the pseudospectral Chevyshev method as a function of $mM$ with $\kappa=5$ (black line), $\kappa=30$ (red line), and $\kappa=100$ (blue line). $\Lambda M^2=0.01$ (top panel), $\Lambda M^2=0.04$ (central panel), and $\Lambda M^2=0.11$ (bottom panel). Left (right) panels for the negative (positive) chirality.} 
\label{XXX}
\end{figure}

{\bf{Anomalous decay rate.}} As we mentioned, the anomalous decay rate behaviour occurs when the longest-lived modes are the ones with higher angular number, while that the existence of a critical scalar field mass indicates that beyond this value the anomalous decay rate behaviour is inverted \cite{Lagos:2020oek,Aragon:2020tvq}. In order to visualize this behaviour, in  Fig. \ref{XXX} we plot $-Im(\omega_{PS})M$  as a function of $mM$, for different values of the cosmological constant $\Lambda M^2=0.01, 0.04$, and $0.11$ and for different values of the parameter $\kappa=5$, $30$, and $100$ by using the pseudospectral Chevyshev method.
We can observe for small values of $\Lambda M^2=0.01$ (top panels) that the anomalous decay rate in the QNFs 
is present, as well as, the existence of a critical mass, where the behaviour is inverted;
however, for $\Lambda M^2=0.04$ (center panels) and for near extremal black hole $\Lambda M
^2=0.11$ (bottom panels), the pseudospectral Chevyshev method predicts that there is not such anomalous behaviour, and also there is not a critical mass for the fermionic field, except for the case  $\Lambda M
^2=0.11$ and positive chirality (right-bottom panel) where we can observe an anomalous behaviour and a small critical mass. So, the existence of anomalous decay rate  depends on the value of $\Lambda M^2$, and the existence of the critical mass depend on $\Lambda M^2$ and the chirality, for the cases considered here.

\clearpage

Now, we shall employ the Wentzel-Kramers-Brillouin (WKB) approach in order to get some analytical insight on the behaviour of the QNFs in the eikonal limit $\kappa \rightarrow \infty$. The WKB approximation was initiated by Mashhoon \cite{Mashhoon} and by Schutz and Iyer \cite{Schutz:1985zz}. Then, Iyer and Will computed the third order correction \cite{Iyer:1986np}, and it was
extended to the sixth order \cite{Konoplya:2003ii}, and up to the 13th order \cite{Matyjasek:2017psv}, see also \cite{Konoplya:2019hlu}. This method has been used to determine the QNFs for asymptotically flat and asymptotically de Sitter black holes. This is due to the WKB method can be used for effective potentials which have the form of potential barriers that approach to a constant value at the horizon and spatial infinity \cite{Konoplya:2011qq}.
The QNMs are determined by the behavior of the effective potential near its maximum value $r^*_{max}$. The Taylor series expansion of the potential around its maximum is given by
\begin{equation}
V(r^*)= V(r^*_{max})+ \sum_{i=2}^{i=\infty} \frac{V^{(i)}}{i!} (r^*-r^*_{max})^{i} \,,
\end{equation}
where
\begin{equation}
V^{(i)}= \frac{d^{i}}{d r^{*i}}V(r^*)|_{r^*=r^*_{max}}\,,
\end{equation}
corresponds to the $i$-th derivative of the potential with respect to $r^*$ evaluated at the maximum of the potential $r^*_{max}$. Using the WKB approximation up to 6th order the QNFs are given by the following expression \cite{Hatsuda:2019eoj}

\begin{eqnarray} \label{principal}
\omega_{PS}^2 &=& V(r^*_{max})  -2 i U \,,
\end{eqnarray}
where
\begin{eqnarray}
\notag U &=&  N\sqrt{-V^{(2)}/2}+\frac{i}{64} \left( -\frac{1}{9}\frac{V^{(3)2}}{V^{(2)2}} (7+60N^2)+\frac{V^{(4)}}{V^{(2)}}(1+4 N^2) \right)  \\
\notag && +\frac{N}{2^{3/2} 288} \Bigg( \frac{5}{24} \frac{V^{(3)4}}{(-V^{(2)})^{9/2}} (77+188N^2) +\frac{3}{4} \frac{V^{(3)2} V^{(4)}}{(-V^{(2)})^{7/2}}(51+100N^2)\\
\notag &&  +\frac{1}{8} \frac{V^{(4)2}}{(-V^{(2)})^{5/2}}(67+68 N^2)+\frac{V^{(3)}V^{(5)}}{(-V^{(2)})^{5/2}}(19+28N^2)+\frac{V^{(6)}}{(-V^{(2)})^{3/2}} (5+4N^2)  \Bigg)\,,
\end{eqnarray}
and $N=n_{PS}+1/2$, with $n_{PS}=0,1,2,\dots$, is the overtone number. We will consider only the effective potential $V_+(r)$ with the plus sign in Eq. (\ref{Pott}) due to $V_-$ yields the same QNFs. Note that in this case the potential also depends on the frequencies, so the evaluation of the QNFs is more difficult; however, our interest is to evaluate the QNFs for large values of $\kappa$, so we 
expand the frequencies as a power series in $\kappa$. In the previous section, we showed in Eq. (\ref{QN}) that in the eikonal limit the leading order term is linear in $\ell$, so we consider the following expansion of the frequency in powers of $\kappa$
\begin{equation}
\omega_{PS} = \omega_1 \kappa + \omega_0+\omega_{-1} \kappa^{-1} + \omega_{-2} \kappa^{-2}\,.
\end{equation}
We find that for large values of $\kappa$, the maximum of the potential is approximately at
\begin{eqnarray}
\notag r_{max} \approx  &&   3M-\frac{\sqrt{3}}{2} M \sqrt{1-9 \Lambda M^2} \kappa^{-1} - \frac{(1-9 \Lambda M^2)(1- 9 M^2 (\Lambda +3 m \omega_1)) mM}{3 \omega_1} \kappa^{-2} \\
\notag && + \frac{M \sqrt{1-9 \Lambda M^2}}{48 \sqrt{3} \omega_1^2} \Bigg( 3 (11-18 \Lambda M^2) \omega_1^2-108 (m M)^2 \omega_1^2 (1-36 \Lambda M^2) + \\
&& 4 m(1-9 \Lambda M^2) \left( 4 \sqrt{3} \omega_0 \sqrt{1-9 \Lambda M^2} + \omega_1 (1-36 \Lambda M^2) \right) \Bigg) \kappa^{-3}  \,,
\end{eqnarray}
and
\begin{eqnarray} \label{coa}
\notag V(r^*_{max}) \approx && \frac{(1-9 \Lambda M^2)}{27 M^2} \kappa^2 - \frac{(1-9 \Lambda M^2)(4m (1- 9M^2 (\Lambda+3 m \omega_1))-3m \omega_1)}{324M^2\omega_1} \\
 && +\frac{(1-9 \Lambda M^2)(6 m (1-9 \Lambda M^2) \omega_0 + \sqrt{3} \sqrt{1-9 \Lambda M^2} \omega_1^2)}{486 M^2 \omega_1 ^2} \kappa^{-1} \,,
\end{eqnarray}
and the second derivative of the effective potential evaluated at $r_{max}^*$ yields
\begin{equation}
V^{(2)}_{r^*_{max}} \approx  -\frac{2 (1-9 \Lambda M^2)^2}{729 M^4} \kappa^2 + \frac{(1-9 \Lambda M^2 )^2 ( 4m (1+ 9 \Lambda M^2 (4-45 \Lambda M^2))-15 \omega_1 +54 \omega_1 M^2(m^2 (4-90 \Lambda M^2)+\Lambda))}{13122 \omega_1 M^4} \,.
\end{equation}

\begin{figure}[h]
\begin{center}
\includegraphics[width=0.46\textwidth]{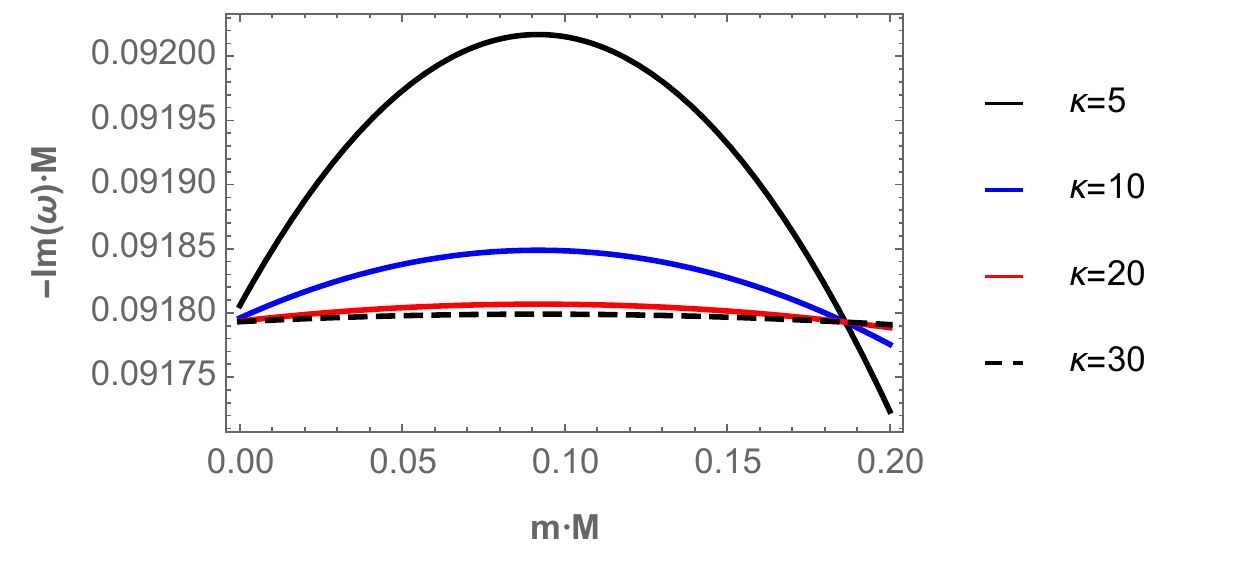}
\includegraphics[width=0.46\textwidth]{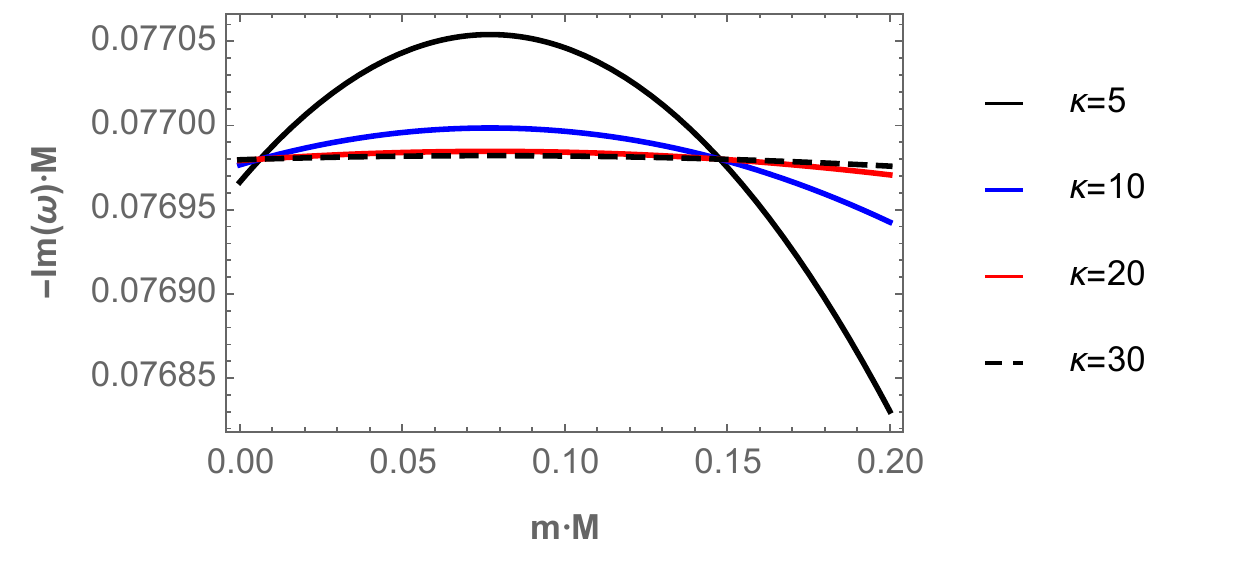}
\includegraphics[width=0.46\textwidth]{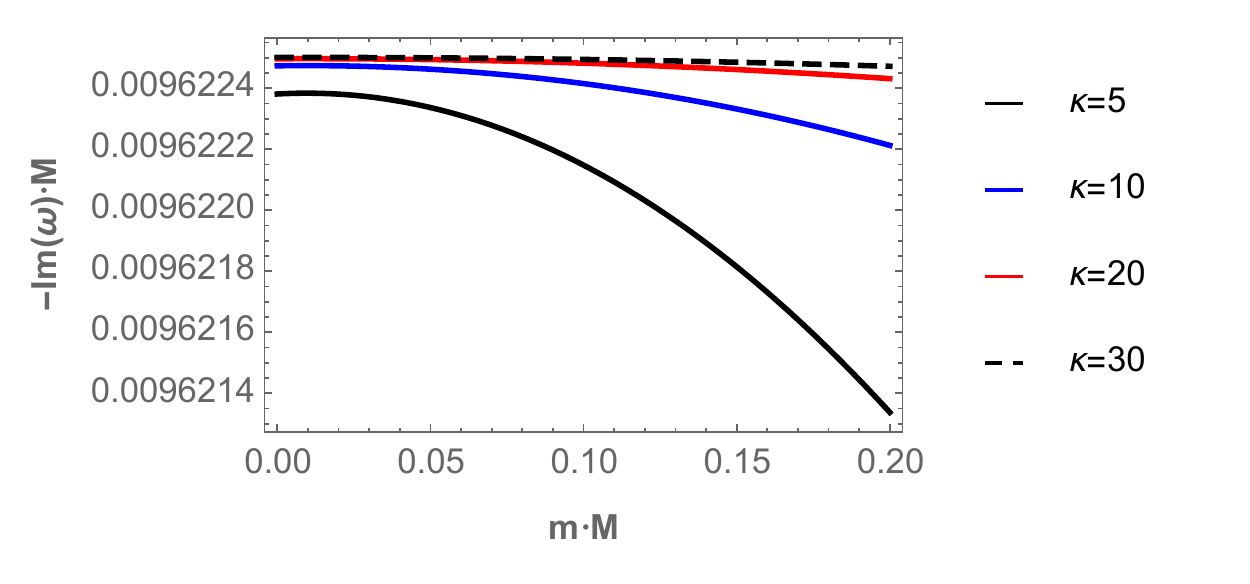}
\end{center}
\caption{The behaviour of $-Im(\omega_{PS}) M$ obtained through the WKB method as a function of $mM$ with $\kappa=5,10,20,30$, and $\Lambda M^2=0.01$ (top left panel), $\Lambda M^2=0.04$ (top right panel), and $\Lambda M^2=0.11$ (bottom panel).} 
\label{WKB}
\end{figure}

The higher derivatives are given by
\begin{eqnarray}
V^{(3)}_{r^*_{max}} \approx && \frac{4(1-9 \Lambda M^2)^3}{6561 M^5} \kappa^2 -\frac{(1-9 \Lambda M^2)^{5/2}}{729 \sqrt{3} M^5} \kappa\,, \\
V^{(4)}_{r^*_{max}} \approx && \frac{16(1-9 \Lambda M^2)^3}{19683 M^6}\kappa^2 +\frac{20 (1- 9 \Lambda M^2)^{7/2}}{19683 \sqrt{3} M^6} \kappa\,,  \\
V^{(5)}_{r^*_{max}} \approx && -\frac{40 (1-9 \Lambda M^2)^4}{59049 M^7} \kappa^2\,, \\
V^{(6)}_{r^*_{max}} \approx && =- \frac{16 (1-9 \Lambda M^2)^4 (4+15 \Lambda M^2)}{177147 M^8} \kappa^2   \,,
\end{eqnarray}
where the leading and subleading terms are important for $V_{r^*_{max}}^{(3)}$ and $V_{r^*_{max}}^{(4)}$, and only the leading terms of the expansion are important for $V_{r^*_{max}}^{(5)}$ and $V_{r^*_{max}}^{(6)}$ in the limit considered. Thus, using these results and $n_{PS}=0$ we obtain $U$ evaluated at $r^*_{max}$

\begin{eqnarray}
 \notag U &=&  (1-9 \Lambda M^2)  \Bigg(  \frac{1}{54 M^2} \kappa - i\frac{65 - 99 \Lambda M^2}{5832 M^2}   \\
\notag && - \Big( \frac{216 m(1-9\Lambda M^2) (1+45 \Lambda M^2)+ 5832 m^2 M^2 (2-45 \Lambda M^2)\omega_1 }{209952 M^2 \omega_1}+ \\
&&  \frac{(169 +216\sqrt{3} i \sqrt{1-9 \Lambda M^2} -  9  \Lambda M^2 (14+1395 \Lambda M^2) ) \omega_1}{209952 M^2 \omega_1} \Big) \kappa^{-1}  \Bigg) \,,
\end{eqnarray}
and 
then replacing
these expansions in Eq. (\ref{principal}) and solving order by order for $\omega_i$ we obtain the following QNFs for $n_{PS}=0$

\begin{eqnarray}
\label{WPS}
\omega_{PS} M &=& \sqrt{1-9 \Lambda M^2} \Bigg( \Big( \frac{1}{3\sqrt{3}} \kappa -\frac{ 11 \sqrt{3} (1-9 \Lambda M^2)  + 324 m M \sqrt{1-9 \Lambda M^2} -972 \sqrt{3} (m M)^2}{1944} \kappa^{-1} \Big) -i\Big(  \frac{1}{6 \sqrt{3}} +  \\
\notag &&   \frac{\sqrt{3} (29-1395 \Lambda M^2)(1-9 \Lambda M^2)+9720 m M (1- 9 \Lambda M^2) \sqrt{1- 9\Lambda M^2 } - 29160 \sqrt{3} (m M)^2 (1-9 \Lambda M^2)}{69984} \kappa^{-2} \Big) \Bigg) \,.
\end{eqnarray}
Notice that for $m=0$ we recover the result ($\ref{QN}$) for $\ell >> 1$. Now, in Fig. \ref{WKB}, we show the behaviour of $-Im(\omega_{PS})M$ given by Eq. (\ref{WPS})  as a function of $mM$, for different values of the cosmological constant $\Lambda M^2=0.01, 0.04$, and $0.11$ and for different values of the parameter $\kappa=5,10,20$, and $30$. We can observe that for small values of $\Lambda M^2=0.01$ the anomalous decay rate of the QNFs of massive fermionic fields in Schwarzschild de Sitter black holes backgrounds is present, in  which the longest-lived modes are the ones with higher angular number,
for a fermionic field mass smaller than
a critical value, while that beyond this value the behaviour is inverted (top left panel). Then, when we increase the value of $\Lambda M^2$ to $0.04$, it is possible to observe (top right panel), the existence of two zones where the longest-lived modes are the ones with smaller angular number, for small and large mass of the fermionic field, and also a central zone where the anomalous behaviour in the decay rate of the QNFs is present. Also, it is possible to observe the existence of two values of critical mass, where the behaviour of the decay rate of the QNF is inverted. Finally, by increasing the value of $\Lambda M^2$ to $0.11$, we can observe (bottom panel) that the anomalous behaviour is not present and the longest-lived modes are the ones with smaller angular number for all the range of fermionic field mass.

\newpage

The critical mass $m_c$ corresponds to the value of the mass for which the term proportional to $\kappa^{-2}$ of $Im(\omega_{PS}) M$ vanishes, thereby
\begin{eqnarray}\label{mc}
 m_c M  \approx &&  \frac{90(1-9\Lambda M^2) \pm \sqrt{30} \sqrt{(299-3825 \Lambda M^2)(1- 9 \Lambda M^2)}}{540\sqrt{3} \sqrt{1-9 \Lambda M^2}} \,.
\end{eqnarray}
The critical mass depends on the values of $\Lambda M^2$. In some cases there is no critical mass, one critical mass or two critical masses, see Fig. \ref{WKB}. In the case of one critical mass, for $m>m_c$, $\omega_{PS}$ increases with $\kappa$; whereas, for $m<m_c$, $-Im(\omega_{PS})$ decreases when $\kappa$ increases. Also, for $\Lambda \rightarrow 0$, we obtain the critical mass $m_c M \approx 0.1975$. On the other hand, when $\Lambda M^2$ increases the critical mass decreases and for values of $\Lambda M^2$ bigger than the critical value $\Lambda_c M^2 = 299 /3825 \approx 0.0782$ there is no critical mass.

\begin{figure}[h]
\begin{center}
\includegraphics[width=0.4\textwidth]{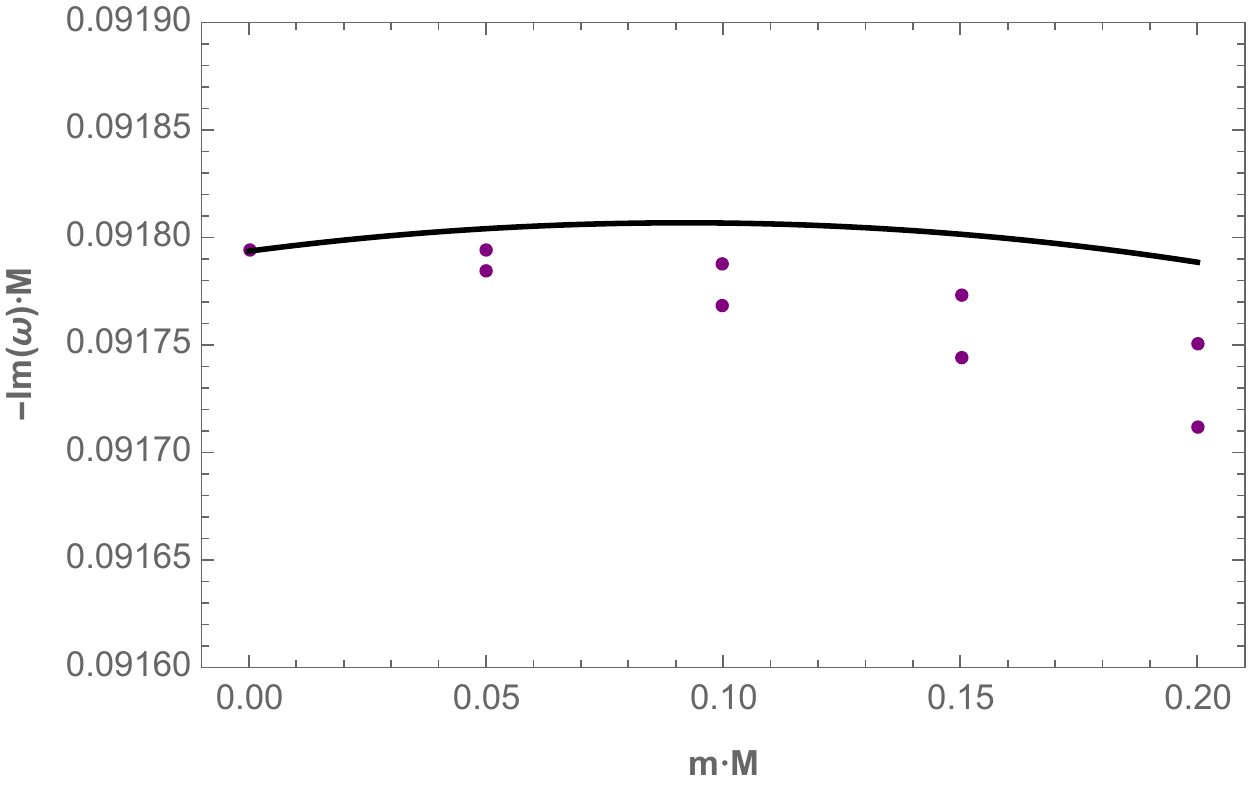}
\end{center}
\caption{The behaviour of $-Im(\omega_{PS})M$ as a function of $mM$, for $\kappa=20$ and $\Lambda M^2=0.01$. Black line is obtained by using the WKB method Eq. (\ref{WPS}), and the purple points are obtained by using the pseudospectral Chevyshev method. }
\label{WvC}
\end{figure}

It is worth mentioning that by using the WKB method it is not possible to observe the fine structure in the spectrum, contrary to the pseudospectral Chevyshev method, where the fine structure appears spontaneously. Moreover, In Fig. \ref{WvC}, we show the behaviour of $-Im(\omega)M$ by using Eq. (\ref{WPS}) (black line), and by using the pseudospectral Chevyshev method (purple points), as a function of $mM$, for $\Lambda M
^2=0.01$, and $\kappa=20$, we can observe the difference in the $-Im(\omega)M$ for both methods. 
We attribute the discrepancy in the value of the critical mass in both methods, WKB and numerical, to the assertion that the WKB approach is not very accurate for the case of massive fermionic field, this is because the WKB method strictly cannot be applied to the massive case, 
due to the effective potential allows 
another local minimum, so that the problem has now three turning points, as was proved in Ref. \cite{Konoplya:2017tvu}. So, the WKB approach fails to give the accurate value of the critical mass.


\subsection{ de Sitter modes}

The de Sitter family consist of the modes that continuously approach to the QNMs of pure de Sitter spacetime when $M \rightarrow 0$. The frequencies of QNMs in pure de Sitter spacetime are given by \cite{Du:2004jt}

\begin{equation}
\omega=-m -i\sqrt{\frac{\Lambda}{3}} \left( 2n_{dS}+ \kappa_++\frac{3}{2}\right)\,, \,\,\,\,\, \text{or} \,\,\,\,\, \omega = m -i\sqrt{\frac{\Lambda}{3}} \left(2n_{dS}+\kappa_+ +\frac{1}{2} \right) \,,
\end{equation}
and
\begin{equation}
\omega=-m -i \sqrt{\frac{\Lambda}{3}} \left( 2n_{dS}- \kappa_-+\frac{1}{2} \right)\,, \,\,\,\,\, \text{or} \,\,\,\,\, \omega = m -i \sqrt{\frac{\Lambda}{3}} \left( 2n_{dS}-\kappa_- -\frac{1}{2} \right) \,,
\end{equation}
where $\kappa_+$ and $\kappa_-$ are positive and negative integers, respectively, and $n_{dS}=0,1,2, \dots$. 

\begin{figure}[h]
\begin{center}
\includegraphics[width=0.5\textwidth]{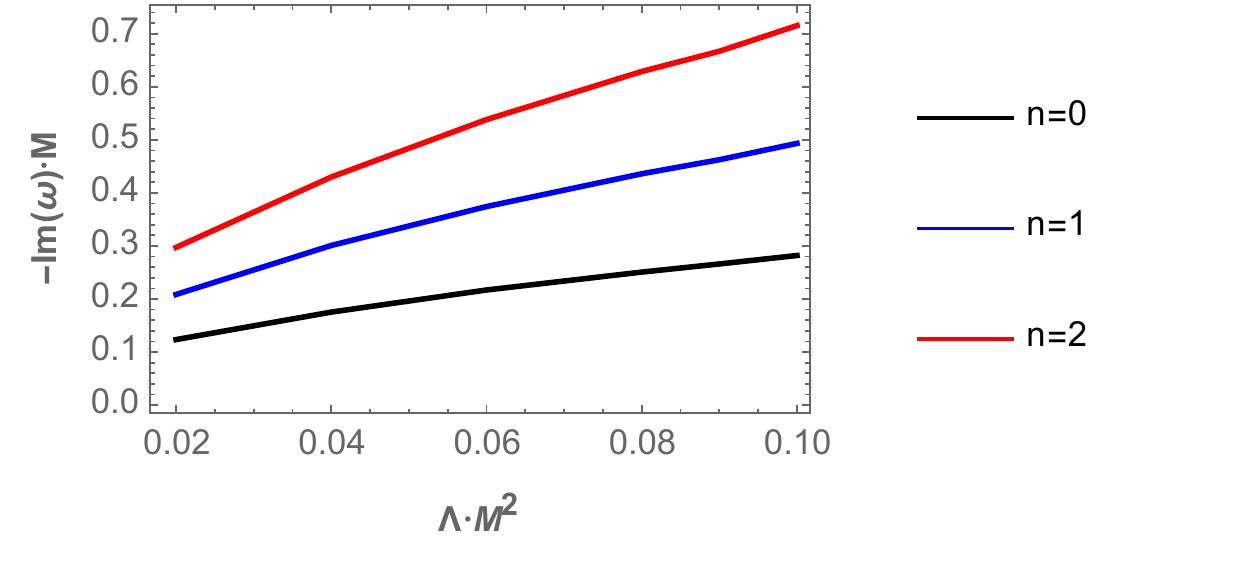}
\end{center}
\caption{The behaviour of $-Im(\omega_{dS}) M$ obtained through the pseudospectral Chevyshev method as a function of $\Lambda M^2$ for massless fermionic fields with $\kappa=1$, and different values of the overtone number $n_{dS}=0,1,$ and $2$.} 
\label{dS}
\end{figure}

Now, in order to visualize the behaviour of the dS modes obtained through the pseudospectral Chevyshev method we plot $-Im(\omega_{dS}) M$ as a function of $\Lambda M^2$ in Fig. \ref{dS} for massless fermionic fields with $\kappa=1$, and different values of the overtone number $n_{dS}=0,1,$ and $2$. First, we can observe that for small values of $\Lambda M^2$, the dS modes tend to the pure de Sitter modes with $\kappa=\kappa_+=-\kappa_-=1$. Also,  the decay rate  increases when $\Lambda M^2$ or the overtone number increases. However, the dS modes are present in a range of value of $\kappa$ and $n_{dS}$, see Table \ref{dST}.  

\begin {table}[ht]
\caption {Quasinormal frequencies $\omega_{dS} M$ for massless fermionic fields obtained through the pseudospectral Chevyshev method for  different values of  $\kappa$ in the background of Schwarzschild-dS black holes with $\Lambda M^2 = 0.04$. }
\label {dST}\centering
\begin {tabular} { | c | c | c | c |}
\hline
$\kappa$ & $n_{dS} = 0 $ & $n_{dS} = 1 $ & $n_{dS} = 2 $ \\\hline
$1$ &
$-0.175313983 i$ &
$-0.300886899 i$ &
$-0.429950083 i$ \\
$2$ &
$-0.289834484 i$ &
$-0.410535743 i$ &
$-0.534933226 i$ \\
$3$ &
$-0.404895443 i$ &
$-0.523977837 i$ &
$-0.645964460 i$ \\
$10$ &
$-1.212651098 i$ &
$-1.329255647 i$ &
$-1.44693078 i $ \\\hline
\end {tabular}
\end {table}

\begin{figure}[h]
\begin{center}
\includegraphics[width=0.4\textwidth]{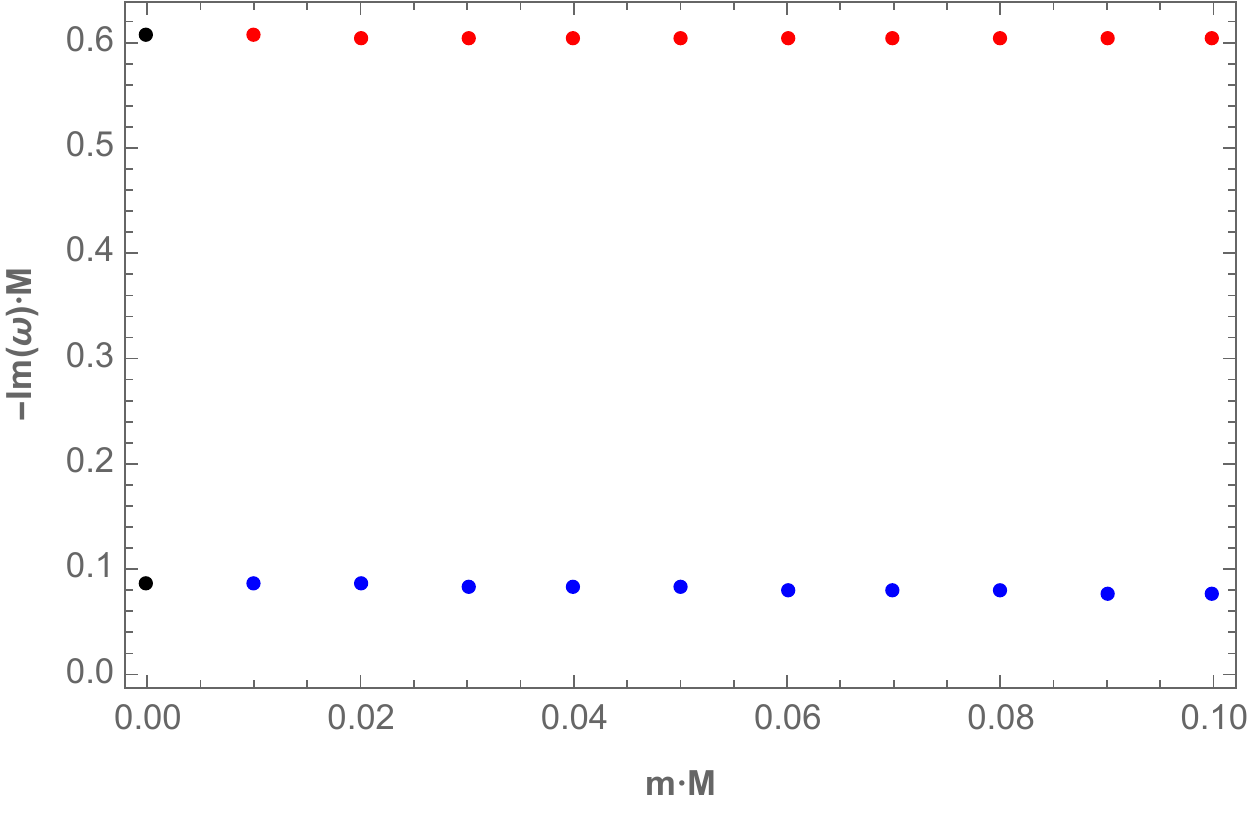}
\includegraphics[width=0.4\textwidth]{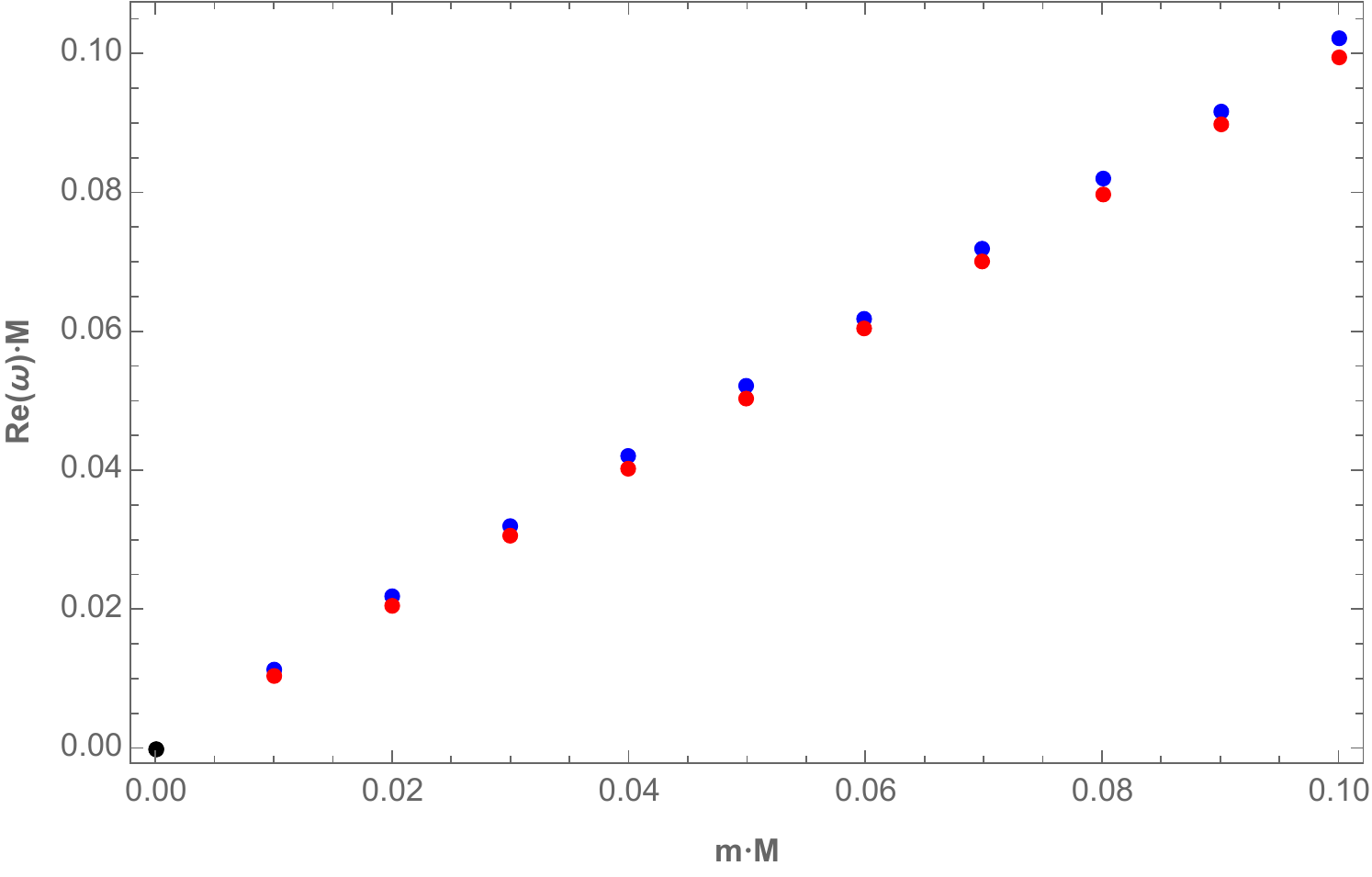}
\end{center}
\caption{The behaviour of $-Im(\omega_{dS})M$ (left panel), and $Re(\omega_{dS})M$ (right panel) obtained through the pseudospectral Chevyshev method as a function of $mM$, for $\kappa=1$ (blue points), and $\kappa=10$ (red points), with $\Lambda M^2=0.01$.}%
\label{ds}
\end{figure}

 Now, in order to observe the behaviour of the dS modes we plot in Fig. \ref{ds} the QNFs as a function of $mM$ for $\kappa=1$ (blue points), and $\kappa=10$ (red points), with $\Lambda M^2=0.01$, and $n_{dS}=0$. The black points indicate a purely imaginary QNFs, that occurs for $mM=0$, then this family acquires a real part. We can observe that the decay rate increases when the angular number increases, and exhibits a smooth behaviour when $mM$ increases (left panel). On the other hand, the frequency of the oscillations increases when $mM$ increases and when the angular number decreases (right panel).       

\newpage

\section{Final remarks}
\label{conclusion}
We considered the propagation of fermionic fields in the background of Schwarzschild de Sitter black holes. Then, we showed the existence of two families of QNMs, one of them corresponds to the photon sphere modes and the other one to the dS modes. Mainly, we showed that it is possible to observe a fine structure in the spectrum, and also an anomalous behaviour for the photon sphere modes, and it is not possible to observe the same for the dS modes. Also, both families present
frequencies with a negative imaginary part, which means that the propagation of fermionic field is stable in this background for the values of the parameters that we considered; however, due to the existence of negative gaps in the potentials the stability is not evident and it  
could be studied  via the time-domain integration of the scalar wave equation as was performed for massless Dirac field on Schwarzschild de Sitter black holes \cite{Konoplya:2020zso}. 

Our analysis was performed by using the pseudospectral Chebyshev method, and the WKB approach. However, while the results for massless fermionic field are similar with a small percentage of error, for massive fermionic field, the result are different and show different behaviours. The difference is due to the assertion that the WKB approach is not very accurate for the case of massive fermionic field, this is because the WKB method strictly cannot be applied to the massive case, due to the effective potential allows another local minimum, so that the problem has now three turning points, as was proved in Ref. \cite{Konoplya:2017tvu}.

Our conclusions are based on pseudospectral Chebyshev method, where the fine structure related to the coupling between the chirality and the mass of the field appears spontaneously in the spectrum, contrary to the WKB where we can not observe the fine structure. We showed that the fine structure is proper of the photon sphere modes, and the separation in the $-Im(\omega_{PS})M$ and $Re(\omega_{PS})M$ between the modes with positive and negative chirality  decreases when the parameter $\kappa$ increases, it increases for higher overtone numbers, and it is more finer when $\Lambda M^2$ increases. 

Also, the decay rate of QNMs of fermionic perturbations show an anomalous behaviour and the presence of a critical fermionic field mass for small values of $\Lambda M^2$, for intermediate values we showed that there is not anomalous behaviour either critical fermionic field mass. However, for higher values, i.e when the black hole becomes near-extremal, we found an anomalous decay rate for the modes associated with a positive chirality and a small fermionic field critical mass. While that for the modes associated with a negative chirality there is not an anomalous behaviour either fermionic field critical mass. Also, we showed that there is a value of $\Lambda M^2$ where the imaginary parts of the modes with positive and negative chirality coincide, before this value the modes with positive chirality are longer lived than the modes with negative chirality, and for higher values of $\Lambda M^2$ the behaviour is inverted, the modes with negative chirality become longer lived.

\acknowledgments

We thank the referee for his/her careful review of the manuscript and his/her valuable comments and suggestions which helped us to improve the manuscript. Y.V. acknowledge support by the Direcci\'on de Investigaci\'on y Desarrollo de la Universidad de La Serena, Grant No. PR18142.

\end{document}